\documentclass{article}
\usepackage[pdftex]{graphicx}
\usepackage{amsmath}
\usepackage{amsbsy}
\usepackage{amsopn}
\usepackage{amsfonts}
\usepackage{amssymb}
\usepackage[square]{natbib}
\usepackage{url}
\usepackage{subfigure}
\usepackage{booktabs}
\usepackage{rotating}
\usepackage[hang,footnotesize,bf]{caption}
\usepackage{setspace}
\usepackage{multirow}
\usepackage{epstopdf}

 \usepackage{lineno}
 \linenumbers*[1]

\usepackage{colortbl}
\arrayrulecolor{black}
\doublerulesepcolor{black}

\newcommand*{\IsDraft}{false}

\newcommand*{\threequarterwidth}{12.5cm}

\newcommand{\be}{\begin{equation}}
\newcommand{\ee}{\end{equation}}
\newcommand{\ba}{\begin{eqnarray}}
\newcommand{\ea}{\end{eqnarray}}

\newcommand*{\model}[1]{\textsc{#1}}

\begin{document}
\title{Retrospective Evaluation of the Five-Year and Ten-Year CSEP-Italy Earthquake Forecasts }

\author{
Maximilian J.\ Werner\textsuperscript{1}\textsuperscript{*},
J. Douglas Zechar\textsuperscript{1,2},
Warner Marzocchi\textsuperscript{3},
Stefan Wiemer\textsuperscript{1},\\
and the CSEP-Italy Working Group\textsuperscript{4}}
\date{\today}
\maketitle

\begin{center}
revised version submitted 22 July 2010 \\
first version submitted 3 March 2010 to the CSEP-Italy special issue of the Annals of Geophysics \\
\end{center}

\vskip0.3cm

\noindent \textsuperscript{1} Swiss Seismological Service, Institute of Geophysics, ETH Zurich, Switzerland. \\
\textsuperscript{2} Lamont-Doherty Earth Observatory of Columbia University, Palisades, USA. \\
\textsuperscript{3} Istituto Nazionale di Geofisica e Vulcanologia, Rome, Italy. \\
\textsuperscript{4} See the Acknowledgements Section.

\noindent\textsuperscript{*} Corresponding author:  \\
\hspace* {4cm} Maximilian J. Werner \\
\hspace* {4cm} Swiss Seismological Service \\
\hspace* {4cm} ETH Zurich\\
\hspace* {4cm} Sonneggstr. 5\\
\hspace* {4cm} 8092 Zurich, Switzerland\\
\hspace* {4cm} {\bf mwerner@sed.ethz.ch}

\vskip2cm

\noindent{\it Online Material}: Additional figures of earthquake forecasts, likelihood ratios, and concentration diagrams.\\
\noindent Available at \url{http://mercalli.ethz.ch/~mwerner/CSEP_ITALY/ESUPP/}

\newpage

\doublespacing

\abstract

On 1 August 2009, the global Collaboratory for the Study of Earthquake Predictability (CSEP) launched a prospective and comparative earthquake predictability experiment in Italy. The goal of the CSEP-Italy experiment is to test earthquake occurrence hypotheses that have been formalized as probabilistic earthquake forecasts over temporal scales that range from days to years. In the first round of forecast submissions, members of the CSEP-Italy Working Group presented eighteen five-year and ten-year earthquake forecasts to the European CSEP Testing Center at ETH Zurich. We considered the twelve time-independent earthquake forecasts among this set and evaluated them with respect to past seismicity data from two Italian earthquake catalogs. In this article, we present the results of tests that measure the consistency of the forecasts with the past observations. Besides being an evaluation of the submitted time-independent forecasts, this exercise provided insight into a number of important issues in predictability experiments with regard to the specification of the forecasts, the performance of the tests, and the trade-off between the robustness of results and experiment duration. We conclude with suggestions for the design of future earthquake predictability experiments. 

\vskip0.5cm




\section{Introduction}
\label{sec:Intro}

On August 1, 2009, a prospective and competitive earthquake predictability experiment began in the region of Italy \citep{Schorlemmer-et-al2010}. The experiment follows the design proposed by the Regional Earthquake Likelihood Model (RELM) working group in California \citep{Field2007, Schorlemmer-et-al2007, SchorlemmerGerstenberger2007, Schorlemmer-et-al2010r} and falls under the global umbrella of the Collaboratory for the Study of Earthquake Predictability (CSEP) \citep{Jordan2006, Zechar-et-al2009}. Eighteen five-year forecasts that express a variety of scientific hypotheses about earthquake occurrence were submitted to the European CSEP Testing Center at ETH Zurich. In this article, we present the results from testing these forecasts retrospectively on seismicity data from two Italian earthquake catalogs. 


The rationale for performing these retrospective tests is as follows: 
\begin{enumerate}
\item [I/] To verify that the submitted forecasts are as intended by the modelers;
\item [II/] To provide a sanity-check of the forecasts before the end of the five-year or ten-year experiments;
\item [III/] To provide feedback to each modeler about the performance of her or his model in retrospective tests and to encourage model improvements;
\item [IV/] To better understand the tests and performance metrics; 
\item [V/] To have worked, pedagogical examples of plausible observations and results; and
\item [VI/] To understand the relation between the duration of predictability experiments and the robustness of the outcomes. 
\end{enumerate}
Nevertheless, retrospective tests also come with significant caveats:
\begin{enumerate}
\item We only evaluated time-independent models; to fairly test time-dependent models on past data would require that the model software be installed at the testing center so that hindcasts could be generated. We identify long-term forecasts from time-dependent models in section \ref{sec:Models} and we did not analyze these forecasts.
\item Past data may be of lower quality than the data used for prospective testing (e.g. greater uncertainties in magnitude and location, or missing aftershocks, potentially with systematic bias). 
\item There are many versions of the past, in the form of several available earthquake catalogs. In an attempt to address this issue, we tested with respect to two catalogs (see section \ref{sec:Data}).
\item All of the forecasts considered here are in some way based on past observations. For example, parameters of the models typically were optimized on part or all of the data against which we retrospectively tested the models. Therefore, positive retrospective test results might simply reveal that a model can adequately fit the data on which it was calibrated, and they might not be indicative of future performance on independent data. A study beyond the scope of this article would be required to decide which of the retrospective data can be regarded as out-of-sample for each model. On the other hand, poor performance of a time-independent forecast in these retrospective experiments indicates that the forecast cannot adequately explain the available data. Therefore, one aim of this article is to identify forecasts of time-independent models that consistently fail in retrospective tests, thereby separating ineffective time-independent models from potentially good models. 
\end{enumerate}

Poor performance of a time-independent forecast might result from one or more of several factors: technical errors (i.e., errors in software implementation), a misunderstanding of the required object to be forecast, calibration on low-quality data, evaluation with low-quality data, statistical type II errors, or incorrect hypotheses of earthquake occurrence. CSEP participants seek to minimize the chances of each of these effects save the final one: that a forecast is rejected because its underlying hypotheses about earthquake occurrence are wrong. 


This article is accompanied by an electronic supplement (available online at \url{http://mercalli.ethz.ch/~mwerner/CSEP_ITALY/ESUPP/}); the reader can find additional figures and a table of information gains that aid in the evaluation of the considered forecasts.

\section{Overview of the Time-Independent Models}
\label{sec:Models}

Each of the forecasts submitted to the five- and ten-year CSEP-Italy experiment can be broadly grouped into one of two classes: those derived from time-independent models, and those that derive from time-dependent models (see Table \ref{tab:models}). The forecasts in the former class are considered to be suitable for any time translation and they depend only on the length of the forecasting time interval (at least over a reasonable time interval where the models are assumed to be time-independent). Therefore, these forecasts can be tested on different target periods. In contrast, the forecasts in the latter group depend on the initial time of the forecast. Because the recipes for calculating the forecasts (i.e. the model software) were not available to us, we could not generate hindcasts from these models that could be meaningfully evaluated. We therefore did not consider time-dependent models in this study. Below, we provide a brief summary of each time-independent model. 

The model \model{Akinci-et-al.Hazgridx} contains the assumption that future earthquakes will occur close in space to locations of historical $m \geq 4$ mainshocks. No tectonic, geological or geodetic information was used to calculate the forecast. The model is based on the method by \citet{Weichert1980} to estimate the seismic rate from declustered earthquake catalogs whose magnitude completeness threshold varies with time. The forecast uses a Gutenberg-Richter law with a uniform b-value. 



\model{Chan-et-al.Hzati} considers a specific bandwidth function to smooth past seismicity and to evaluate the spatial seismicity density of earthquakes. The model smoothes both spatial locations and magnitudes. The smoothing procedure is applied to a coarse seismotectonic zonation based on large-scale geological structure. The expected rate of earthquakes is obtained from the average historical seismicity rate. 



Each Asperity Likelihood Model (ALM)--\model{Gulia-Wiemer.ALM}, \model{Gulia-Wiemer.HALM},\\
\model{Schorlemmer-Wiemer.ALM}--hypothesizes that small-scale spatial variations in the b-value of the Gutenberg-Richter relationship play a central role in forecasting future seismicity \citep{WiemerSchorlemmer2007}. The physical basis of the model is the concept that the local b-value is inversely proportional to applied shear stress. Thus low b-values ($b < 0.7$) are thought to characterize the locked patches of faults (asperities) from which future mainshocks are more likely to be generated, whereas high b-values ($b > 1.1$), found for example in creeping sections of faults, suggest a lower probability of large events. The b-value variability is mapped on a grid. The local a and b-values in the forecasts \model{Gulia-Wiemer.ALM} and \model{Gulia-Wiemer.HALM} were obtained from the observed rates of declustered earthquakes between 1981 and 2009, using Reasenberg's declustering method \citep{Reasenberg1985} and the Entire-Magnitude-Range method for completeness estimation by \citet{WoessnerWiemer2005}. In the \model{Gulia-Wiemer.HALM} model (Hybrid ALM), a "hybrid" between a grid-based and a zoning model, the Italian territory is divided into distinct regions depending on the main tectonic regime and the local b-value variability is thus mapped using independent b-values for each tectonic zone. In the \model{Schorlemmer-Wiemer.ALM} model, derived from the original ALM  \citep{WiemerSchorlemmer2007}, the authors decluster the input catalog (2005-2009) for $m \geq  2$ using the method by \citet{GardnerKnopoff1974} and smooth the node-wise rates of the declustered catalog with a Gaussian filter. Completeness values for each node are taken from the analysis by \citet{Schorlemmer-et-al2010b} using the probability-based magnitude of completeness method. The resulting forecast is calibrated to the observed average number of events with $m \geq 4.95$.



The \model{Meletti-et-al.MPS04} model \citep[][\url{http://zonesismiche.mi.ingv.it}]{MPS2004} is the reference model for seismic hazard in Italy. \model{Meletti-et-al.MPS04} derives from the standard approach to probabilistic seismic hazard assessment of \citet{Cornell1968}, in which a Poisson process is assumed. The model distributes the seismicity in a seismotectonic zonation and it considers the historical catalog using, through a logic tree structure, two different ways (historical and statistical) to estimate its completeness. The models also assumes that each zone is characterized by its own Gutenberg-Richter law with varying truncation.

The Relative Intensity (RI) model (\model{Nanjo-et-al.RI}) is a pattern recognition model based on the main assumption that future large earthquakes tend to occur where the seismic activity had a specific pattern (usually a higher seismicity) in the past. In its first version, the RI code was "alarm-based," i.e., the code made a binary statement about the occurrence of earthquakes. For the CSEP-Italy experiment, the code was modified to estimate the expected number of earthquakes in a specific time-space-magnitude bin. 

The models \model{Werner-et-al.CSI} and  \model{Werner-et-al.Hybrid} are based on smoothed seismicity. Future earthquakes are assumed to occur with higher probability in areas where past earthquakes have occurred. Locations of past mainshocks are smoothed using an adaptive power-law kernel, i.e little in regions of dense seismicity, more in sparse regions. The degree of smoothing is optimized via retrospective tests. The magnitude of each earthquake is independently distributed according to a tapered Gutenberg-Richter distribution with corner magnitude $8.0$. The model uses small magnitude $m \geq 2.95$ quakes, whenever trustworthy, to better forecast future large events. The two forecasts \model{Werner-et-al.CSI} and \model{Werner-et-al.Hybrid} were obtained by calibrating the model on two different earthquake catalogs. 


The forecasts \model{Zechar-Jordan.CPTI}, \model{Zechar-Jordan.CSI} and \model{Zechar-Jordan.Hybrid} are derived from the Simple Smoothed Seismicity (Triple-S) model, which is based on Gaussian smoothing of past seismicity.  Past epicenters make a smoothed contribution to an earthquake density estimation, where the epicenters are smoothed using a fixed lengthscale $\sigma$; $\sigma$ is optimized by minimizing the average area skill score misfit function in a retrospective experiment \citep{ZecharJordan2010}. The density map is scaled to match the average historical rate of seismicity. The two forecasts  \model{Zechar-Jordan.CPTI} and \model{Zechar-Jordan.CSI} were optimized for two different catalogs, while \model{Zechar-Jordan.Hybrid} is a hybrid forecast.

\section{Specification of CSEP-Italy Forecasts}
\label{sec:Forecasts}
We use the term ``seismicity model" to mean a system of hypotheses and inferences that is presented as a mathematical, numerical and simplified description of the process of seismicity. A ``seismicity forecast" is a statement about some observable aspect of seismicity that derives from a seismicity model. In the context of the CSEP-Italy experiment, a seismicity forecast is a set of estimates of the expected number of future earthquakes in each bin, where bins are specified by intervals of location, time and magnitude within the multi-dimensional testing volume \citep[see also][]{Schorlemmer-et-al2007}. More precisely, the CSEP-Italy participants agreed (within the official ``Rules of the Game" document) to provide a numerical estimate of the likelihood distribution of observing any number of earthquakes within each bin. Moreover, this discrete distribution, which specifies the probability of observing zero, one, two, etc earthquakes in a bin, is given by a Poisson distribution (defined below in section \ref{sec:NBD}) which is uniquely defined by the expected number of earthquakes. Each bin's distribution is assumed independent of the distribution in other bins, and the observed number of earthquakes in a given bin is compared with the forecast of that bin.

\section{Data Used For Retrospective Testing}
\label{sec:Data}

For prospective tests of the submitted forecasts, the Italian Seismic Bulletin ({\it Bollettino Sismico Italiano}, BSI) recorded by INGV will be used \citep[see][]{Schorlemmer-et-al2010}. We did not use the BSI for a retrospective evaluation of forecasts because it is only available in its current form since April 2005. Instead, we used two alternative Italian earthquake catalogs provided by the INGV, which were also provided as a tool for the modelers for model learning and calibration: the {\it Catalogo Parametrico dei Terremoti Italiani} (Parametric Catalog of Italian Earthquakes, CPTI08) \citep{CPTI08} and the {\it Catalogo del la Sismicit`a Italiana} (Catalog of Italian Seismicity, CSI 1.1) \citep{Castello-et-al2007, Chiarabba-et-al2005}. \citet{Schorlemmer-et-al2010} discuss the catalogs in detail, we only provide a brief overview. Both data sets are available for download from \url{http://www.cseptesting.org/regions/italy}.

\subsection{The CSI 1.1 Catalog 1981--2002}
\label{sec:CSI}

The CSI catalog spans the time period from 1981 until 2002 and reports local magnitudes, in agreement with the BSI magnitudes that will be used during the prospective evaluation of forecasts. \citet{Schorlemmer-et-al2010} found a clear change in earthquake numbers per year in 1984 due to the numerous network changes in the early 1980s and therefore recommend using the CSI data from 1 July 1984 onwards. For the retrospective evaluation, we selected earthquakes with local magnitudes $M_L \geq 4.95$ from 1985 until the end of 2002. To mimic the durations of the prospective experiments, we selected three non-overlapping five-year periods (1998-2002, 1993-1997, 1988-1992). To test the robustness of the results, we also used the entire 18-year span of reliable data from 1985 until 2002. We selected shocks as test data if they occurred within the CSEP-Italy testing region \citep[see][]{Schorlemmer-et-al2010}.

\subsection{The CPTI08 Catalog 1901--2006}
\label{sec:CPTI}

The CPTI catalog covers the period from 1901 until 2006 and is based on both instrumental and historical observations (for details, see \citet{Schorlemmer-et-al2010}). The catalog lists moment magnitudes that were estimated either from macroseismic data or calculated using a linear regression relationship between surface wave, body wave or local magnitudes. Because the prospective experiment will use local magnitudes, we converted the moment magnitudes to local magnitudes using the same regression equation that was used to convert the original local magnitudes to moment magnitudes for the creation of the CPTI catalog \citep{MPS2004, Schorlemmer-et-al2010}:
\begin{equation}
M_L=1.231 (M_W - 1.145) \ .
\label{eq:ML}
\end{equation}

\citet{Schorlemmer-et-al2010} estimated a conservative completeness magnitude of $M_L=4.5$, so that one could justify using the entire period from 1901 until 2006 for the retrospective evaluation. However, we focused mainly on the data since the 1950s because it seems to be of higher quality \citep{Schorlemmer-et-al2010}. We divided the period into non-overlapping ten-year periods to mimic the duration of the prospective experiment, but we also evaluated the forecasts on a 57-year time span from 1950 until 2006 and on the 106-year period from 1901 until 2006. As for the CSI catalog, we only selected shocks within the testing region. Some quakes, mostly during the early part of the CPTI catalog, were not assigned depths. We included these earthquakes as observations within the testing region because it is very unlikely that they were deeper than 30 km (see also \citet{Schorlemmer-et-al2010}).

\subsection{The Distribution of the Number of Earthquakes}
\label{sec:NBD}

In this section, we consider the distribution of the number of observed events in the five- and ten-year periods relevant for the Italian forecasts. Analysis of this empirical distribution can test the assumption (made by all time-independent forecasts) that the Poisson distribution approximates well the observed variation in the number of events in each cell and in the entire testing region. (CSEP-Italy participants decided to forecast all earthquakes and not only so-called mainshocks -- see section \ref{sec:NBDf}).

The Poisson distribution is defined by its discrete probability mass function: 
\begin{equation}
p (n | \lambda) = \lambda^n \frac{\exp (-\lambda)}{n!} \ ,
\label{eq:Poi}
\end{equation}
where $n=0,1,2,...,$ and $\lambda$ is the rate parameter, the only parameter needed to define the distribution. The expected value and the variance $\sigma^2_{POI}$ of the Poisson distribution are both equal to $\lambda$. 

Because the span of time over which the CSI catalog is reliable is so short, we used the CPTI catalog for the seismicity rate analysis. The sample variance of the distribution of the number of observed earthquakes in the CPTI catalog over non-overlapping five-year periods from 1907 until 2006 (inclusive) is $\sigma^2_{5yr} = 23.73$, while the sample mean is  $\mu_{5yr} = 8.55$. For non-overlapping ten-year periods of the CPTI catalog, the sample variance is $\sigma^2_{10yr} = 64.54$, while the sample mean is $\mu_{10yr} = 17.10 $. Because the sample variance is so much larger than the sample mean on the five- and ten-year timescales, it is clear that the seismicity rate varies more widely than expected by a Poisson distribution.  

\begin{figure}
\centering
\includegraphics[draft=\IsDraft,width=\threequarterwidth,keepaspectratio=true,clip]{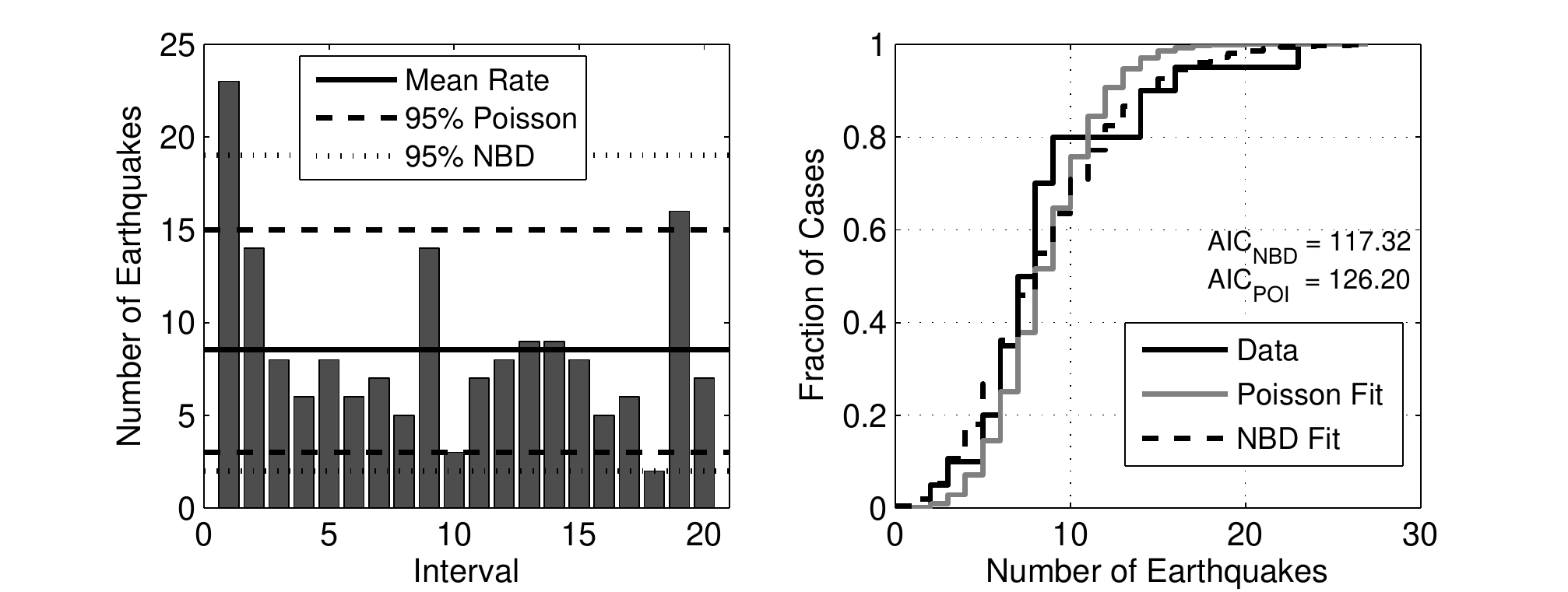}
\caption{\label{fig:NumDist}Left: Observed number of earthquakes in twenty non-overlapping five-year intervals in the CPTI catalog from 1907 until 2006 (inclusive) (bars), mean number of observed events (solid line) along with two-sided $95\%$ confidence bounds from the Poisson distribution (\ref{eq:Poi}) and the negative binomial distribution (NBD) (\ref{eq:NBD}). Right: Empirical cumulative distribution function (solid black line), along with fits to the data using a Poisson distribution (solid grey line) and an NBD (dashed black line). Also shown are the Akaike Information Criterion (AIC) values of the fitted distributions (\ref{eq:AIC}). 
}
\end{figure}

Figure \ref{fig:NumDist} shows the number of observed earthquakes in each of the twenty non-overlapping five-year intervals, along with the empirical cumulative distribution function. The Poisson distribution with $\lambda=\mu_{5yr} = 8.55$ and its $95\%$ confidence bounds are also shown. One should expect that one in twenty data points falls outside the $95\%$ confidence interval, but we observe four, and one of these lies outside the $99.99\%$ quantile. 

We compared the goodness of fit of the Poisson distribution with that of a negative binomial distribution (NBD), motivated by studies that suggest its use based on empirical and theoretical considerations \citep{VereJones1970, Kagan1973, JacksonKagan1999, Kagan2009a, Schorlemmer-et-al2010r, Werner-et-al2009b}.The discrete negative binomial probability mass function is
\begin{equation}
p(n| \tau, \nu)= \frac{\Gamma(\tau+n)}{\Gamma(\tau) n!} \ \nu^{\tau} (1-\nu)^n \ ,
\label{eq:NBD}
\end{equation}
where $n=0,1,2,...,$ $\Gamma$ is the gamma function, $0 \leq \nu \leq 1$, and $\tau > 0$. The average of the NBD is given by
\begin{equation}
\mu_{NBD}= \tau \frac{1-\nu}{\nu} \ ,
\label{eq:NBDAvg}
\end{equation}
while the variance is given by
\begin{equation}
\sigma^2_{NBD}=\tau \frac{1-\nu}{\nu^2} \ ,
\label{eq:NBDVar}
\end{equation}
implying that $\sigma^2_{NBD} \geq \sigma^2_{POI}$. \citet{Kagan2009a} discusses different parameterizations of the NBD. For simplicity, we used the above parameterization and maximum likelihood parameter value estimation. We found $\tau_{5yr} = 6.49$ and $\nu_{5yr} = 0.43$, with $95\%$ confidence bounds given by $[-0.39,13.37]$ and $[0.17,0.70]$, respectively. The large uncertainties reflect the small sample size of twenty. For the ten-year intervals, we estimated $\tau_{10yr} = 9.24$ and $\nu_{10yr} = 0.35$, with $95\%$ confidence bounds given by $[-2.74,21.22]$ and $[0.05,0.65]$, respectively. Figure \ref{fig:NumDist} shows the $95\%$ confidence bounds of the fitted NBD in the number of observed events (left panel), and the NBD cumulative distribution function (right panel).

Because the NBD is characterized by two parameters while the Poisson has only one, we used the Akaike Information Criterion (AIC) \citep{Akaike1974} to compare the fits:
\begin{equation}
AIC = 2 p-2 \log (L)  \ ,
\label{eq:AIC}
\end{equation}
where $L$ is the likelihood of the data given the fitted distribution and $p$ is the number of free parameters. For the five-year and ten-year intervals, the NBD fit the data better than the Poisson distribution, despite the penalty for the extra parameter: for the five-year intervals, $AIC_{NBD}=117.32$ and $AIC_{POI}=126.20$, while for the ten-year intervals, $AIC_{NBD}=70.05$ and $AIC_{POI}=77.56$. To test the robustness of the better fit of the NBD over the Poisson distribution, we also checked the distribution of the number of events in one-year, two-year and three-year intervals of both catalogs. In all cases, the NBD fit better than the Poisson distribution, despite the penalty for an extra parameter. 

\subsection{Implications for the CSEP-Italy Experiment}
\label{sec:NBDf}

Several previous studies showed that the distribution of the number of earthquakes in any finite time period is not well approximated by a Poisson distribution and is better fit by an NBD \citep{Kagan1973, JacksonKagan1999, Schorlemmer-et-al2010r, Werner-et-al2009b} or a heavy-tailed distribution \citep{SaichevSornette2006c}. The implications for the CSEP-Italy experiment, and indeed for all CSEP experiments to date, are important. 

The only time-independent point process is the Poisson process \citep{DaleyVereJones2003}. Therefore, a non-Poissonian distribution of the number of earthquakes in a finite time-period implies that, if a point process can model earthquakes well, this process must be time-dependent (although there might be other, non-point-process classes of models that are time-independent and generate non-Poissonian distributions). Therefore, the Poisson point process representation is inadequate, even on five- or ten-year timescales for large $m \geq 4.95$ earthquakes in Italy, because the rate variability of time-independent Poisson forecasts is too small, and they will fail more often than expected. As a result, the agreement of CSEP-Italy participants to use a Poisson distribution should be viewed as problematic for time-independent models because no Poisson distribution that their model could produce will ever pass the tests with the expected $95\%$ confidence rate. On the other hand, time-dependent models with varying rate  might generate an NBD over a longer time period (the empirical NBD can even be used as a constraint on the model).

Solutions to this problem have been discussed previously. The traditional approach has been to filter the data via declustering (deletion) of so-called aftershocks (as done, for instance, in the RELM mainshock experiment \citep{Field2007, Schorlemmer-et-al2007}). However, the term ``aftershock" is model-dependent and can only be applied retrospectively. A more objective  approach is to forecast all earthquakes, allowing for time-dependence and non-Poissonian variability. In theory, each model could predict its own distribution for each space-time-magnitude bin \citep{WernerSornette2008a}, and future predictability experiments should consider allowing modelers to provide such a comprehensive forecast (see also section \ref{sec:Disc}). 

A third, ad-hoc solution \citep[see][]{Werner-et-al2009b} is more practical for time-independent models in the current context. Based on an empirical estimate of the observed variability of past earthquake numbers, one can reinterpret the original Poisson forecasts of time-independent models to create forecasts that are characterized by an NBD. One can perform all tests (defined below in section \ref{sec:Tests}) using the original Poisson forecasts, and repeat the tests with so-called NBD forecasts. 

We created NBD forecasts for the total number of observed events by using each forecast's mean and an imposed variance identical for all models, which we estimated either directly from the CPTI catalog or from extrapolation assuming that the observed number of events are uncorrelated. Appendix A describes the process in detail. Because the resulting NBD forecasts are tested on the same data that were used to estimate the variance, one should expect that the NBD forecasts perform well. The broader NBD results in less specificity, but also fewer unforeseen observations. We will re-examine this ad-hoc solution in the discussion in section \ref{sec:Disc}. 

\section{Tests}
\label{sec:Tests}

To follow the agreed-upon rules of the prospective CSEP-Italy experiment, we used the statistical tests proposed for the RELM experiment and more recent ones that have been implemented within CSEP \citep{Schorlemmer-et-al2007, Schorlemmer-et-al2010r, Zechar-et-al2010}. These include: (i) the N(umber)-test, based on the consistency between the total number of observed and expected earthquakes; (ii) the L(ikelihood)-test, based on the consistency between the observed and expected joint log-likelihood score of the forecast; (iii) the S(pace)-test, based on the consistency between the observed and expected joint log-likelihood score of the spatial distribution of earthquakes; and (iv) the M(agnitude)-test, based on the consistency between the observed and expected joint log-likelihood score of the magnitude distribution of earthquakes. 

The L-test proposed by  \citet{Schorlemmer-et-al2007} is a relatively uninformative test, because the expected likelihood score is influenced by both the entropy of a model and the expected number of earthquakes. As the expected number increases, the expected likelihood score decreases. Therefore, a model that overpredicts the number of earthquakes will tend to underpredict the likelihood score. Because the L-test is one-sided, i.e. a forecast is not rejected if the observed likelihood score is underpredicted, models that overpredict the number of earthquakes might pass the L-test trivially \citep[for a concrete example, see][p. 1190-1191]{Zechar-et-al2010}. As a remedy, we additionally used a conditional L-test \citep{Werner-et-al2009b}, in which the observed likelihood score is compared with expected likelihood scores conditional on the number of observed quakes. In contrast to the S or M-tests, the conditional L-test assesses the joint space-magnitude forecast, but it overcomes the sensitive dependence of the expected likelihood scores on the number of expected events. 

\section{Results}
\label{sec:Results}

\subsection{Testing Five-Year Forecasts on the CSI Catalog}
\label{sec:ResCSIa}

\begin{figure}[ht!]
\centering
\includegraphics[width=\textwidth,keepaspectratio=true,clip]{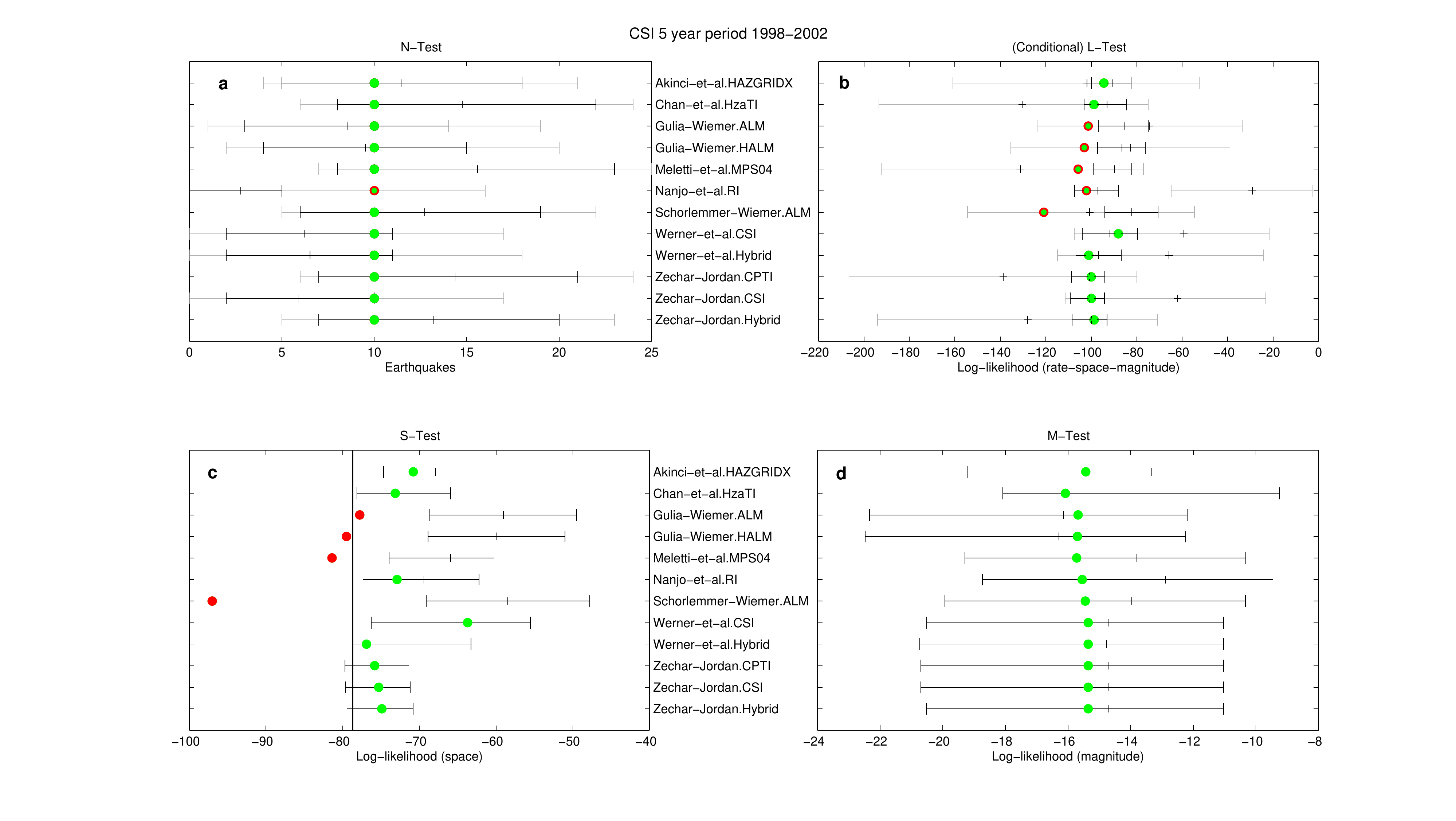}
\caption{\label{fig:CSI5a}Results of the (a) N-test, (b) unconditional and conditional L-tests, (c) S-test and (d) M-test of the 5-year time-independent forecasts using the 5-year target period from 1998 to 2002 of data from the CSI earthquake catalog. Red and green symbols indicate rejected and passed forecasts, respectively. In (a), green symbols with red edges indicate that the Poisson forecast was rejected while the NBD forecast was passed. In (b), green symbols with red edges indicate that only one of the two L-tests was passed. Black crosses: (a) expected number of earthquakes, (b) expected unconditional or conditional log-likelihood score, (c) expected spatial log-likelihood score, (d) expected magnitude log-likelihood score, assuming the forecast is correct. Black bars: $95\%$ confidence bounds of the model forecast assuming a Poisson distribution. In (a), grey bars denote $95\%$ confidence bounds of the model forecast assuming a negative binomial distribution. In (b), black (grey) bars denote $95\%$ confidence bounds of the conditional (unconditional) likelihood score. Vertical lines in S-test figures indicate the likelihood score of a spatially uniform model.
}
\end{figure}

In Figure \ref{fig:CSI5a}, we show the results of the N, L, S, and M-tests applied to the time-independent forecasts for the most recent five-year target period from 1998-2003 of the CSI catalog. We discuss each of the test results below. As a summary of all the results we present here and below, Tables \ref{tab:sumCSI} and \ref{tab:sumCPTI} list all the tests that the forecasts fail for each of the considered target periods of the CSI and CPTI catalog, respectively. 

\subsubsection{N-Test Results}

The N-test results in Figure \ref{fig:CSI5a}a) show that only one forecast (\model{Nanjo-et-al.RI}) can be rejected assuming Poisson confidence bounds because significantly more earthquakes were observed than expected. Using NBD uncertainties, none of the forecasts can be rejected, because the confidence bounds are wider (typically by several earthquakes on both sides). 

\subsubsection{L-Test Results}

In Figure \ref{fig:CSI5a}b), we show the results of the unconditional and the conditional L-tests applied to the original (Poisson) forecasts. We did not try to apply NBD uncertainty to the rate forecasts in each space-magnitude bin, and therefore did not simulate likelihood values based on an NBD forecast. 

Only one forecast fails the unconditional L-test, while four fail the conditional L-test. The confidence bounds of the unconditional L-test are much larger because the number of simulated earthquakes is allowed to vary, thereby increasing the spread of the simulated likelihood scores. The impact of the expected number of earthquakes on the expected unconditional likelihood score is particularly visible for the forecasts  \model{Meletti-et-al.MPS04} and \model{Nanjo-et-al.RI}. The forecast \model{Meletti-et-al.MPS04} expects more earthquakes than were observed during this period (although not significantly more) and therefore also expects a likelihood score that is lower than observed. Moreover, the additional variability due to the increased number of events broadens the confidence bounds and the model thus passes the L-test. However, the the forecast fails the conditional L-test, because, given the number of observed earthquakes, the observed likelihood score is too small to be consistent with the forecast. Meanwhile, the forecast \model{Nanjo-et-al.RI} underpredicts the number of quakes (assuming Poisson variability) and therefore overpredicts the likelihood score and fails the unconditional L-test. However, conditional on the number of observed earthquakes, the observed likelihood score is consistent with the forecast. 

To summarize, the conditional L-test reveals information that is separate from the N-test results and presents a stricter evaluation of the forecasts. In the remainder of this article, we will only consider the more informative conditional L-test results. From the results of the 1998-2002 target period, we can conclude that the joint distribution of the locations and magnitudes of the observed earthquakes are inconsistent with the group of ALM forecasts and the forecast \model{Meletti-et-al.MPS04}. 

\subsubsection{Reference Forecast From a ``Model of Most Information"}
\label{sec:perf}

To quantify the ability of the present time-independent forecasts to accurately predict the locations and magnitudes of the observed earthquakes, one can calculate the likelihood score of an ideal earthquake forecast (what might be called a successful prediction of the observed earthquakes -- naturally with the benefit of hindsight -- or a forecast from a ``model of most information", as opposed to the ``model of least information"  \citep{Evison1999} discussed next). For instance, working within the constraints of a Poisson distribution of events in each bin, one could calculate the likelihood score of a forecast that assigns an expected rate in each space-magnitude bin that is equal to the number of observed shocks within that bin. If at most one earthquake occurs per bin, the observed log-likelihood score of such a perfect forecast is the negative number of observed events. The score is only slightly smaller if more than one event occurs in a given bin. In Figure \ref{fig:CSI5a}b), the observed likelihood scores of the forecasts are evidently ``far" from the score of a perfect forecast, which would roughly equal to $-10$. The typical scores of the forecasts lie in the region of $-100$, which implies that the likelihood of the data under the perfect forecast is about $10^{39}$ more likely than under a typical CSEP-Italy forecast. The information gain per earthquake \citep{HarteVereJones2005} of the perfect forecast over a typical forecast is on the order of $10^4$. 

These numbers help quantify the difference between a perfect ``prediction" within the current CSEP experiment design and a typical probabilistic earthquake forecast. One might imagine tracking this index of earthquake predictability to quantify the progress of the community of earthquake forecasters towards better models. However, the primary goal of CSEP's experiments is to test and evaluate hypotheses about earthquake occurrence, and the observed degree of predictability is sufficient to carry out this endeavor. 

\subsubsection{Reference Forecast From a ``Model of Least Information"}

One could equally construct a forecast from a ``model of least information" \citep{Evison1999}, often called the null hypothesis, which might be based on a uniform spatial distribution, a total expected rate equal to the observed mean over a period prior to the target period, and a Gutenberg-Richter magnitude distribution with b-value equal to one. Because several models already assume that (i) magnitudes are identically and independently distributed according to the Gutenberg-Richter magnitude distribution and (ii) the total expected rate is equal to the mean number of observed shocks, the only real difference between these models and an uninformative forecast lies in the spatial distribution. We therefore included the likelihood score of a spatially uniform forecast only in the S-test results. In Table S1 of the electronic supplement, we additionally provide the information gains per earthquake \citep{KaganKnopoff1977, HarteVereJones2005} of each spatial forecast over a spatially uniform forecast for all the considered target periods.

\subsubsection{S-Test and M-test Results}

The S-test and M-test results, shown in Figures \ref{fig:CSI5a}c) and d), suggest that the weakness of the group of ALM forecasts and the forecast \model{Meletti-et-al.MPS04} lies in forecasting the spatial distribution of earthquakes: all four forecasts fail the S-test with very small p-values, while all models pass the M-test. Additionally, the forecasts \model{Gulia-Wiemer.HALM},  \model{Meletti-et-al.MPS04} and \model{Schorlemmer-Wiemer.ALM} obtain scores that are lower than the score of a uniform model of least information. 

In the case of the ALM group of forecasts, the low spatial likelihood scores leading to the S-test failures have a common origin. In roughly one half of all spatial bins, the three forecasts expect an extremely small constant number of earthquakes per spatial bin, indicating that a constant background rate was set in these cells. The forecasts \model{Gulia-Wiemer.ALM} and \model{Gulia-Wiemer.HALM}  expect on the order of $10^{-8}$ earthquakes in each spatial background bin, while the forecast \model{Schorlemmer-Wiemer.ALM} expects an even smaller $10^{-15}$ earthquakes per bin. Accordingly, the probability of observing one earthquake in these bins is of the same order of magnitude. However, earthquakes do occur in some of these bins, and their occurrences in such low-probability (background) bins cause very low likelihood scores. Because these losses against a normalized uniform forecast, which expects roughly $10^{-3}$ earthquakes per bin to sum to the $10$ observed quakes, are not compensated by equal or greater gains from earthquakes in regions where the forecasts are higher, the forecasts obtain extremely small spatial likelihood scores and fail the S-test. 

During the 1998-2002 period, the forecasts \model{Gulia-Wiemer.ALM} and \model{Gulia-Wiemer.HALM} fail the S-test because of one $M_L5.4$ earthquake, located offshore north of Sicily at $39.06^o$N and $15.02^o$E, which occurred in such a background rate bin. Similarly, the forecast \model{Schorlemmer-Wiemer.ALM} fails the S-test because of a $M_L5.1$ earthquake at $37.93^o$N and $17.55^o$E on the south-eastern boundary of the testing region. Apart from two other events, the remaining seven earthquakes during this target period occurred in cells where the ALM forecasts expected more earthquakes than the uniform forecast. However, the gains achieved for these earthquakes do not compensate the losses incurred from the event in the background bins. 

The distribution of rates of the forecast \model{Meletti-et-al.MPS04} shows the existence of a similar background rate, although it is larger ($10^{-4}$ earthquakes per bin) than the background rates of the ALM forecasts. The occurrence of an earthquake in a background bin can therefore be more easily compensated by gains achieved from other earthquakes. However, during the 1998-2002 period, five earthquakes occurred in such background bins, and the losses were not be masked by the gains. These five earthquakes include all four offshore earthquakes during this period (including the two events that caused the ALM forecasts to fail), along with one additional shock of magnitude $M_L5.3$ at $46.697^o$N and $11.07^o$E in northern Italy. 

\subsubsection{Results from other Five-Year Target Periods of the CSI Catalog}
\label{sec:ResCSI}

\begin{figure}[ht!]
\centering
\includegraphics[width=\textwidth,keepaspectratio=true,clip]{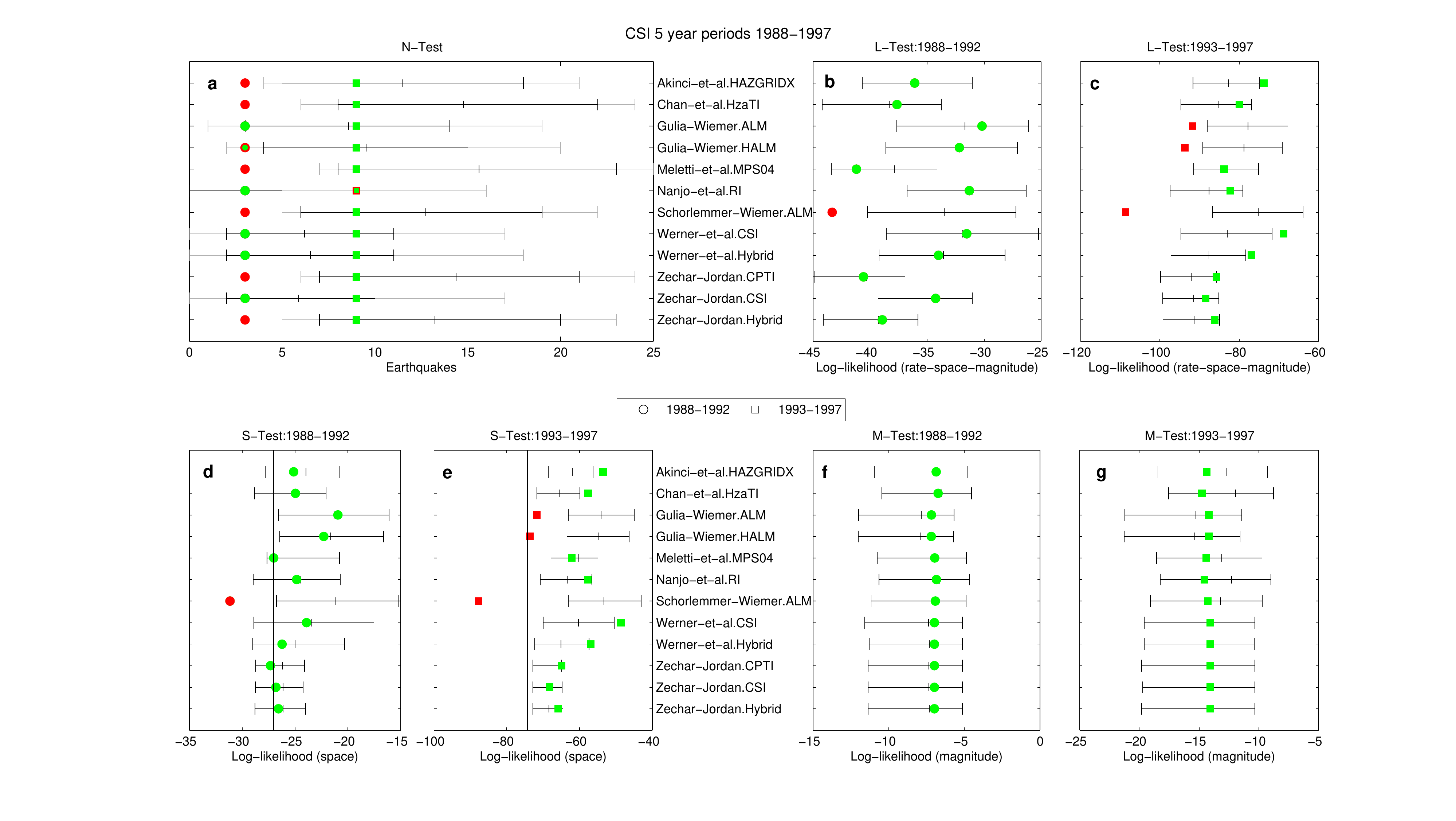}
\caption{\label{fig:CSI5}Results of the (a) N-test, (b-c) conditional L-test, (d-e) S-test and (f-g) M-test of the 5-year time-independent forecasts using two separate 5-year target periods of data from the CSI earthquake catalog. Red and green symbols indicate rejected and passed forecasts, respectively. Green symbols with red edges indicate that the Poisson forecast was rejected while the NBD forecast was passed.  Black crosses: (a) expected number of earthquakes, (b) expected conditional log-likelihood score, (c) expected spatial log-likelihood score, (d) expected magnitude log-likelihood score, assuming the forecast is correct. Black bars: $95\%$ confidence bounds of the model forecast assuming a Poisson distribution. In (a), grey bars denote $95\%$ confidence bounds of the model forecast assuming a negative binomial distribution. Vertical lines in S-test figures indicate the likelihood score of a spatially uniform model.
}
\end{figure}

In Figure \ref{fig:CSI5}, we show the results of two further, separate five-year target periods from the CSI catalog: 1988-1992 (circles) and 1993-1997 (squares). In combination with Figure \ref{fig:CSI5a}, this provides insight into the variability of the five-year test results due to natural fluctuations of seismicity. 

During 1988-1992, only three target earthquakes occurred. Although this number is small, it falls within the $95\%$ confidence bounds of historical fluctuations (see Figure \ref{fig:NumDist}). Six forecasts are rejected by the N-test because they overpredict the number of observed events. These forecasts are: \model{Akinci-et-al.Hazgridx}, \model{Chan-et-al.HzaTI}, \model{Meletti-et-al.MPS04}, \model{Schorlemmer-Wiemer.ALM},  \model{Zechar-Jordan.CPTI}, and \model{Zechar-Jordan.Hybrid}. As results from longer target periods below confirm, this group consistently overpredicts the total rate. The modelers of the forecasts  \model{Akinci-et-al.Hazgridx}, \model{Chan-et-al.HzaTI}, \model{Meletti-et-al.MPS04}, \model{Schorlemmer-Wiemer.ALM},  \model{Zechar-Jordan.CPTI}, and \model{Zechar-Jordan.Hybrid} indicated to us that they calibrated their models on the moment magnitude scale rather than the local magnitude scale used for prospective testing, leading to an overprediction of the number of earthquakes with local magnitude $M_L \geq 4.95$. This error in the calibration complicates the interpretation of the N-test results for this group of models.

As before, we observe differences in the results from the NBD and Poisson N-tests. During 1988-1992, the forecast \model{Gulia-Wiemer.HALM} is rejected by the N-test assuming Poisson confidence bounds, but the more realistic NBD uncertainties allow the forecast to pass. Similarly, the forecast \model{Nanjo-et-al.RI} fails the Poisson N-test but passes the NBD N-test during 1993-1997. 

The conditional L-test results indicate that in the case of the forecast \model{Schorlemmer-Wiemer.ALM}, the three earthquakes during 1988-1992 suffice to reject the model. Results from the 1993-1997 period again show rejections of the ALM group of forecasts. However, in contrast to the 1998-2002 period, the forecast \model{Meletti-et-al.MPS04} passes both periods. Results from longer target periods, presented below, are necessary to judge this forecast conclusively. 

The combined S and M-test results again locate the source of the ALM rejections in the spatial dimension of the forecast. Moreover, \model{Schorlemmer-Wiemer.ALM} continues to perform worse than a uniform model during both target periods. During the 1993-1997 target period, the forecasts fail because of a $M_L5.8$ earthquake in 1994 at $39.398^o$N and $15.21^o$E offshore to the north of Sicily, which occurred in a background bin. The large resulting likelihood loss cannot be compensated by the gains achieved from the other eight observed earthquakes. During the 1988-1992 target period, the forecasts \model{Gulia-Wiemer.ALM} and \model{Gulia-Wiemer.HALM} pass the S-test, but the forecast \model{Schorlemmer-Wiemer.ALM} receives a low likelihood score because of an uncompensated likelihood loss due to a $M_L5.4$ earthquake in 1990 that occurred in a low-probability (but not background) bin at $37.33^o$N and $15.24^o$E offshore and east of Mount Etna. Additionally, the forecast \model{Zechar-Jordan.CPTI} scored marginally less than a uniform forecast, although the score is consistent with the forecast's expectation. 

The M-test results thus far, and for all but the longest of the target periods considered below, are not very informative: no rejections occur. The  individual model distributions are very similar, indicating that the differences between the predicted magnitude distributions are small. The differences between the observed likelihood scores are equally small. 

To summarize, some of the test results vary with the considered five-year target period, while others are robust. \model{Schorlemmer-Wiemer.ALM} consistently shows poor performance in the spatial forecast, while the other two ALM forecasts are rejected in two of three target periods. \model{Meletti-et-al.MPS04} fails the conditional L and S tests during one of three five-year target periods. 

\subsection{Testing Ten-Year Forecasts on the CPTI Catalog}

\begin{sidewaysfigure}[ht!]
\centering
\includegraphics[width=\textwidth,keepaspectratio=true,clip]{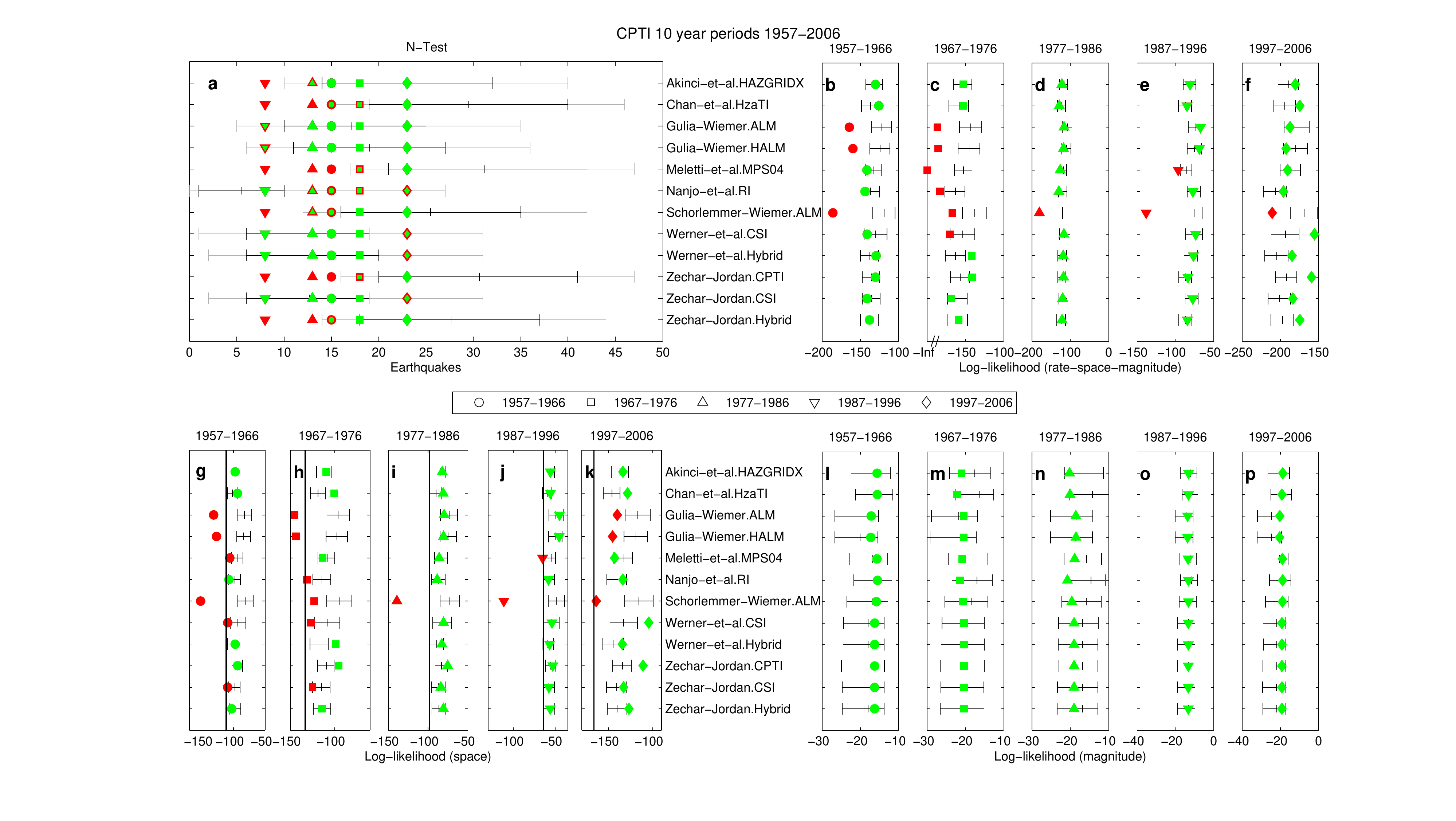}
\caption{\label{fig:CPTI10}Results of the (a) N-test, (b-f) conditional L-test, (g-k) S-test and (l-p) M-test of the 10-year time-independent forecasts using five separate 10-year target periods of data from the CPTI earthquake catalog. Red and green symbols indicate rejected and passed forecasts, respectively. Green symbols with red edges indicate that the Poisson forecast was rejected while the NBD forecast was passed.  Black crosses: (a) expected number of earthquakes, (b-f) expected conditional log-likelihood score, (g-k) expected spatial log-likelihood score, (l-p) expected magnitude log-likelihood score, assuming the forecast is correct. Black bars: $95\%$ confidence bounds of the model forecast assuming a Poisson distribution. In (a), grey bars denote $95\%$ confidence bounds of the model forecast assuming a negative binomial distribution. Vertical lines in S-test figures indicate the likelihood score of a spatially uniform model.
}
\end{sidewaysfigure}

In Figure \ref{fig:CPTI10}, we summarize the results of the N, conditional L, S and M-tests for the time-independent models and five non-overlapping ten-year target periods of the CPTI catalog. These results mimic the prospective ten-year experiment and help gauge the variability of the results. The online material that accompanies this article (available at \url{http://mercalli.ethz.ch/~mwerner/CSEP_ITALY/ESUPP/}) provides additional figures of the forecasts, maps of their likelihood ratios against a uniform forecast, and concentration diagrams \citep{RongJackson2002, Kagan2009} for the entire CPTI data set from 1901 until 2006. Because the figures are based on the longest target period, which we consider explicitly in section \ref{sec:long}, they include all earthquakes observed during the ten-year target periods and provide an informative visual presentation of the results. 

\subsubsection{N-Test Results}

The N-test results are shown in panel a) of Figure  \ref{fig:CPTI10}. The numbers of observed shocks during the five ten-year periods from were 15, 18, 13, 8 and 23. For the remainder of this article, we do not discuss the N-test results from the group of models that were wrongly calibrated on the moment magnitude scale (see section \ref{sec:ResCSI}). Of the remaining six forecasts, none could forecast all five observations within the $95\%$ confidence bounds of the Poisson distribution. Five forecasts -- \model{Gulia-Wiemer.ALM}, \model{Gulia-Wiemer.HALM}, \model{Werner-et-al.CSI}, \model{Werner-et-al.Hybrid} and \model{Zechar-Jordan.CSI} -- are rejected only during one of the five periods when assuming Poisson confidence bounds and cannot be rejected at all when considering confidence bounds based on an NBD. 

The forecast \model{Nanjo-et-al.RI} expects far fewer shocks than the other forecasts and consistently underpredicts the number of earthquakes. Assuming the original Poisson variability in the number of shocks, the forecast is rejected during four of the five target periods. However, the forecast cannot be rejected at all if NBD confidence bounds are used. 

\subsubsection{Conditional L-Test Results}
\label{sec:cpti10L}

The conditional L-test results are displayed in panels b) to f) of Figure \ref{fig:CPTI10}. The only robust result is the continued failure of the  forecast \model{Schorlemmer-Wiemer.ALM}. The forecasts \model{Gulia-Wiemer.ALM} and \model{Gulia-Wiemer.HALM} fail the test during two periods, while \model{Nanjo-et-al.RI} and \model{Werner-et-al.CSI} are both rejected during 1967-1976. Reasons for these rejections are discussed in the context of the S and M-test results below. 

The forecast \model{Meletti-et-al.MPS04} obtains an observed joint-log-likelihood score of negative infinity during the target period 1967-1976. This score results from the fact that one earthquake occurred in a space-magnitude bin in which the forecasted rate was zero.  A zero forecast is equivalent to saying that target earthquakes are impossible in this bin, and if an event does occur in this bin, the forecast is automatically rejected. The earthquake in question, the 1968 Belice earthquake, occurred on 15 January 1968 in western Sicily at 37.76$^o$N and 12.98$^o$E with a magnitude $M_L6.39$ and caused several hundred fatalities. According to the forecast, however, earthquakes larger than $M_L=6.25$ are impossible in this spatial bin because the forecasted rates in the magnitude bins are non-zero only for magnitudes up to $M_L=6.25$. The forecast's rejection implies that the maximum magnitude set for this location was too small; the discrepancy might be due to the wrong magnitude calibration reported above and/or indicate that the maximum magnitude may require a modification in this area. (The forecast \model{Meletti-et-al.MPS04} does not fail the S-test because the forecast in this particular spatial cell is non-zero when summed over the individual magnitude bins.)

\subsubsection{S-Test and M-test Results}

In panels g) through k) of Figure \ref{fig:CPTI10} the S-test results are shown. Five (spatial) forecasts cannot be rejected by the S-test during any of the five target periods: \model{Akinci-et-al.Hazgridx}, \model{Chan-et-al.HzaTI}, \model{Werner-et-al.Hybrid}, \model{Zechar-Jordan.CPTI} and \model{Zechar-Jordan.Hybrid}. 

The two forecasts \model{Werner-et-al.CSI} and \model{Zechar-Jordan.CSI}, which were optimized on the CSI catalog, both fare well during the target periods that are also or at least partially covered by the CSI catalog, i.e. from 1981 onwards. However, the two forecasts are rejected during the two earliest target periods, which can be considered as out-of-sample tests for these two forecasts. During the 1957-1966 period, the forecasts fail to predict several diffuse earthquakes in northern Italy and two offshore earthquakes between the Ligurian coast and Corsica. The 1967-1976 period contains the 1968 western Sicily earthquake sequence (including the above-mentioned $M_L=6.39$ Belice earthquake), which occurs in spatial cells with low expected rates. Evidently, the CSI catalog contains little seismicity in these regions from which the models could have anticipated the occurrence of these earthquakes. 

Interestingly, the forecast \model{Nanjo-et-al.RI}, which was also calibrated on CSI data, only fails during the 1967-1977 period (again due to the western Sicily sequence in 1968) but passes during the 1957-1966 interval. The model employs a relatively coarse grid to forecast earthquakes (see Figure S6 of the electronic supplement), and this characteristic helped forecast the offshore quakes north of Corsica better than the \model{Werner-et-al.CSI} and \model{Zechar-Jordan.CSI} forecasts. 

The three ALM-based forecasts continue to forecast poorly the spatial distribution of observed earthquakes. During the 1957-1966 target period, the two above-mentioned earthquakes north of Corsica and a shock in northern Italy occur in background bins of all three ALM forecasts, leading to their S-test failures. During the 1967-1976 target periods, the \model{Gulia-Wiemer.ALM} and \model{Gulia-Wiemer.HALM} forecasts fail because of three earthquakes in background bins: two shocks occurred as part of the 1968 western Sicily earthquake sequence and one in central Italy at 44.81$^o$N and 10.35$^o$E. While none of these events (nor any others) occur in background bins of the forecast \model{Schorlemmer-Wiemer.ALM} during this period, two earthquakes of the 1968 western Sicily sequence, as well as an earthquake at 41.65$^o$N and 15.73$^o$E, do incur unexpectedly low likelihood scores, resulting in the S-test rejection. In fact, \model{Schorlemmer-Wiemer.ALM} fails all considered ten-year target periods. Whenever the spatial likelihood score falls below a uniform forecast, at least one earthquake occurred in a so-called background cell. 

The forecast \model{Meletti-et-al.MPS04} is rejected twice by the S-test. During the period 1957-1966, the forecast fails because of the two recurring offshore earthquakes north of Corsica in July 1963 and because of two earthquakes in northeastern Italy, all of which occurred in background bins. During 1987-1996, three earthquakes occurred in background bins: (i) an offshore earthquake on April 26, 1988, at $42.21^o$N and $16.66^o$E;  (ii) an $M_L=5.43$ aftershock of the Potenza, southern Italy, earthquake of May 5, 1990; and (iii) an $M_L=5.54$ offshore earthquake on December 13, 1990, east of Mount Etna in the Sea of Sicily. 

\subsection{Test Results from Longer Periods}
\label{sec:long}

The long-term forecasts submitted for the CSEP-Italy experiment were calculated for five-year and ten-year periods. Because the forecasts are time-independent and characterized by Poisson uncertainty, one can test suitably scaled versions of the forecasts on longer time periods: 18 years (the duration of the reliable part of the entire CSI catalog, from 1985 through 2002), 57 years (the duration of the most reliable part of the CPTI catalog, from 1950 through 2006), and 106 years (the entire CPTI catalog). In this section, we  present the results of testing these scaled forecasts. The online material presents further figures of the forecasts, likelihood ratios and concentration diagrams based on the 106-year target period.  

\begin{figure}[ht]
\centering
\includegraphics[width=\textwidth,keepaspectratio=true,clip]{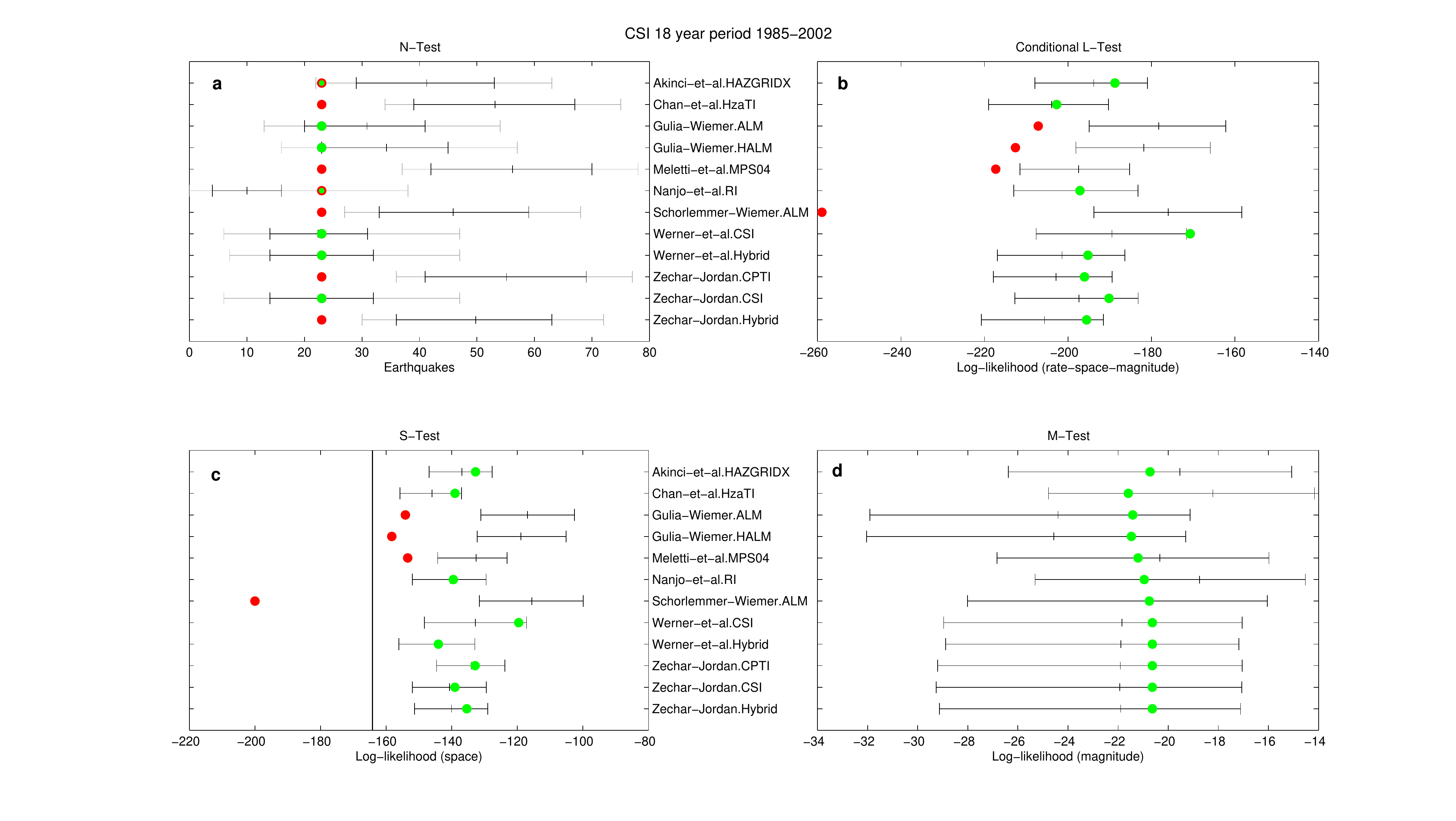}
\caption{\label{fig:CSI18}Results of the (a) N-test, (b) conditional L-test, (c) S-test and (d) M-test of the scaled 10-year time-independent forecasts using the 18-year target period from 1985 to 2002 of data from the CSI earthquake catalog. Red and green symbols indicate rejected and passed forecasts, respectively. Green symbols with red edges indicate that the Poisson forecast was rejected while the NBD forecast was passed.  Black crosses: (a) expected number of earthquakes, (b) expected conditional log-likelihood score, (c) expected spatial log-likelihood score, (d) expected magnitude log-likelihood score, assuming the forecast is correct. Black bars: $95\%$ confidence bounds of the model forecast assuming a Poisson distribution. In (a), grey bars denote $95\%$ confidence bounds of the model forecast assuming a negative binomial distribution. Vertical lines in S-test figures indicate the likelihood score of a spatially uniform model.}
\end{figure}

The test results of the 18-year period from 1985 to 2002 of the CSI catalog are shown in Figure  \ref{fig:CSI18}. Twenty-three earthquakes occurred during this period. The N-test results reveal the same features already observed previously: a group of models overpredicts the number of earthquakes (\model{Akinci-et-al.Hazgridx}, \model{Chan-et-al.HzaTI}, \model{Meletti-et-al.MPS04}, \model{Schorlemmer-Wiemer.ALM}, \model{Zechar-Jordan.CPTI}, \model{Zechar-Jordan.Hybrid}). While the confidence bounds of the negative binomial distribution remain substantially wider than the bounds based on the Poisson distribution, there are only two forecasts for which the test results are ambiguous (\model{Akinci-et-al.Hazgridx} and \model{Nanjo-et-al.RI}). The ALM forecasts and the \model{Meletti-et-al.MPS04} forecast fail the conditional L-test and the S-test, with \model{Schorlemmer-Wiemer.ALM} scoring less than a uniform spatial forecast. The failures are due to the earthquakes we discussed previously that occur either in background bins or in locations with low expected rates. 

\begin{figure}[ht]
\centering
\includegraphics[width=\textwidth,keepaspectratio=true,clip]{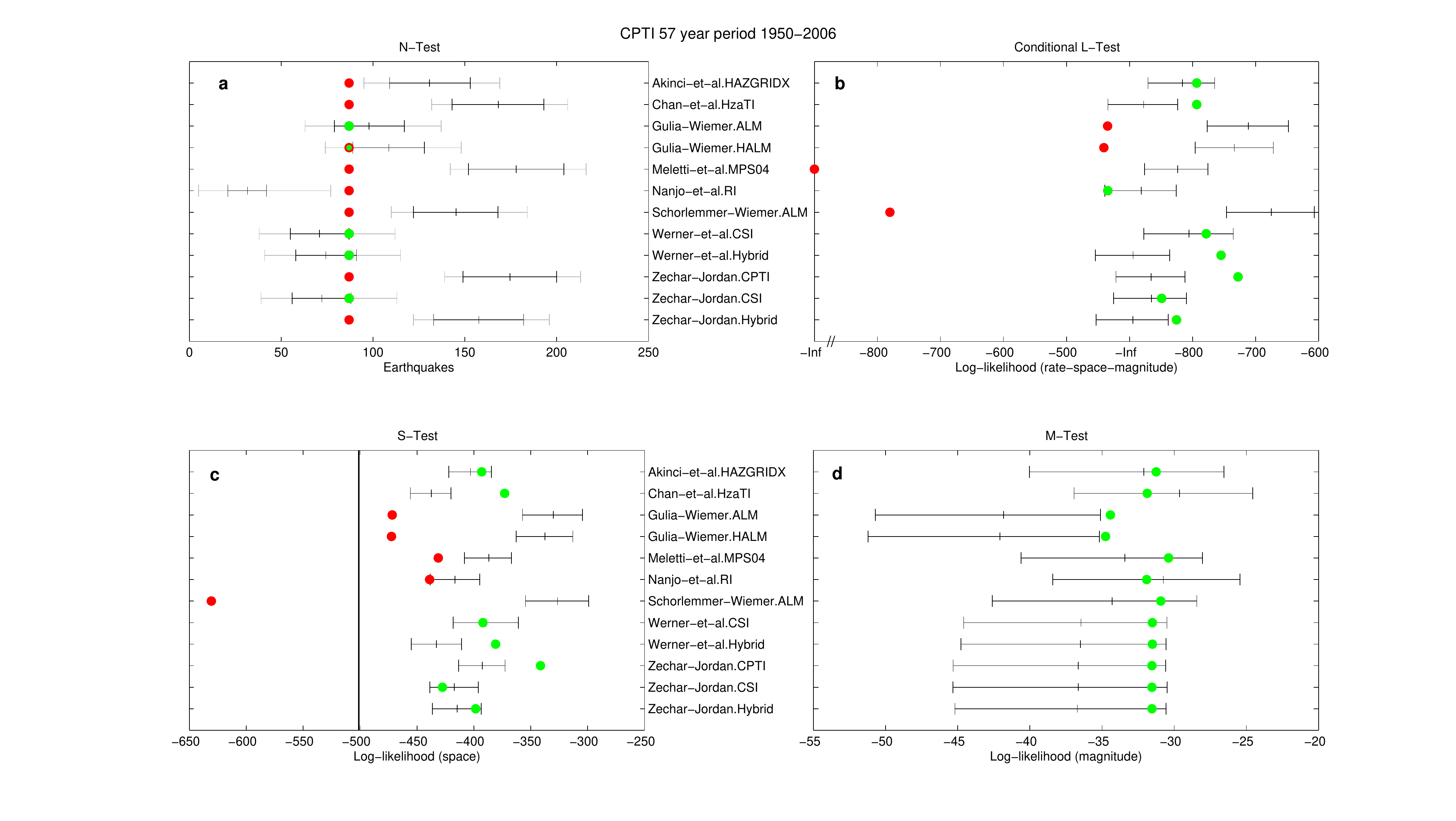}
\caption{\label{fig:CPTI57}Results of the (a) N-test, (b) conditional L-test, (c) S-test and (d) M-test of the scaled 10-year time-independent forecasts using the 57-year target period from 1950 to 2006 of data from the CPTI earthquake catalog. Red and green symbols indicate rejected and passed forecasts, respectively. Green symbols with red edges indicate that the Poisson forecast was rejected while the NBD forecast was passed.  Black crosses: (a) expected number of earthquakes, (b) expected conditional log-likelihood score, (c) expected spatial log-likelihood score, (d) expected magnitude log-likelihood score, assuming the forecast is correct. Black bars: $95\%$ confidence bounds of the model forecast assuming a Poisson distribution. In (a), grey bars denote $95\%$ confidence bounds of the model forecast assuming a negative binomial distribution. Vertical lines in S-test figures indicate the likelihood score of a spatially uniform model.}
\end{figure}

Increasing the duration of the retrospective tests to the 57 most recent years of the CPTI catalog (1950-2007) yields 83 earthquakes and leads to similar results but with greater statistical significance (Figure \ref{fig:CPTI57}). In addition to the rejections mentioned in the preceding paragraph, the N-test now unequivocally rejects the forecasts \model{Akinci-et-al.Hazgridx} and \model{Nanjo-et-al.RI}, even when the confidence bounds of an NBD are considered. The conditional L-test rejects the forecast \model{Meletti-et-al.MPS04} because of a likelihood score of negative infinity (discussed in section \ref{sec:cpti10L}). The S-test results show that the forecast \model{Nanjo-et-al.RI} can be rejected in addition to the ALM forecasts and \model{Meletti-et-al.MPS04}. No forecasts can be rejected by the M-test, despite 57 years of data. 

\begin{figure}[ht]
\centering
\includegraphics[width=\textwidth,keepaspectratio=true,clip]{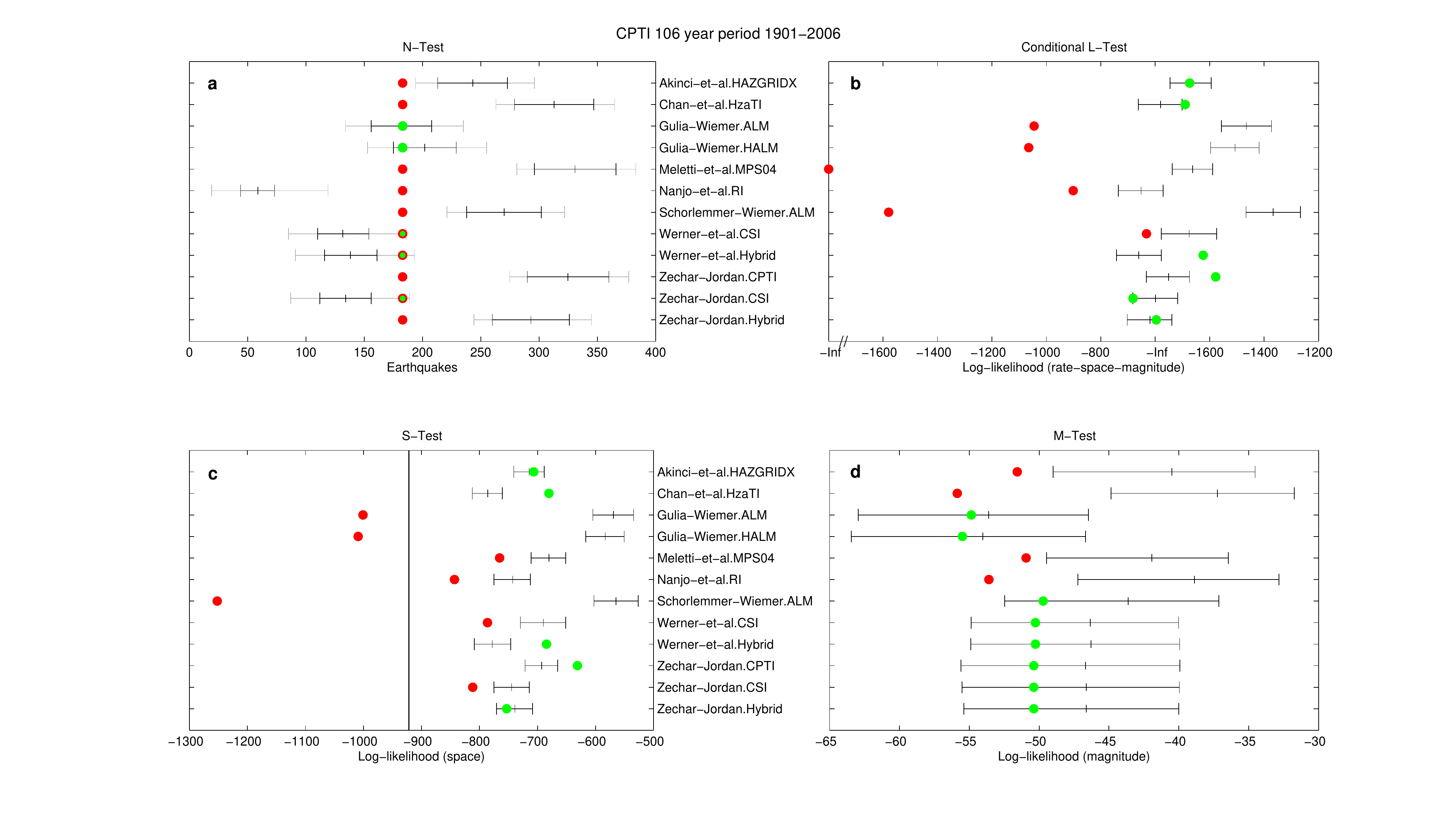}
\caption{\label{fig:CPTI106}Results of the (a) N-test, (b) conditional L-test, (c) S-test and (d) M-test of the scaled 10-year time-independent forecasts using the 106-year target period from 1901 to 2006 of data from the CPTI earthquake catalog. Red and green symbols indicate rejected and passed forecasts, respectively. Green symbols with red edges indicate that the Poisson forecast was rejected while the NBD forecast was passed.  Black crosses: (a) expected number of earthquakes, (b) expected conditional log-likelihood score, (c) expected spatial log-likelihood score, (d) expected magnitude log-likelihood score, assuming the forecast is correct. Black bars: $95\%$ confidence bounds of the model forecast assuming a Poisson distribution. In (a), grey bars denote $95\%$ confidence bounds of the model forecast assuming a negative binomial distribution. Vertical lines in S-test figures indicate the likelihood score of a spatially uniform model.}
\end{figure}

The longest period over which we evaluated the scaled forecasts was 106 years, spanning the full duration of the CPTI08 catalog and containing 183 earthquakes (Figure \ref{fig:CPTI106}, see the online material for maps of the forecasts, likelihood ratios and concentration diagrams). The N-test results now show a clear separation between the group of forecasts that consistently overpredict, the forecast \model{Nanjo-et-al.RI}, which underpredicts, and the forecasts that cannot be rejected by assuming confidence bounds based on either a Poisson or a negative binomial distribution. Application of the conditional L-test additionally rejects the forecasts \model{Nanjo-et-al.RI} and \model{Werner-et-al.CSI}, while the S-test now also fails \model{Werner-et-al.CSI} and \model{Zechar-Jordan.CSI}. 

\begin{figure}[ht]
\centering
\includegraphics[width=\textwidth,keepaspectratio=true,clip]{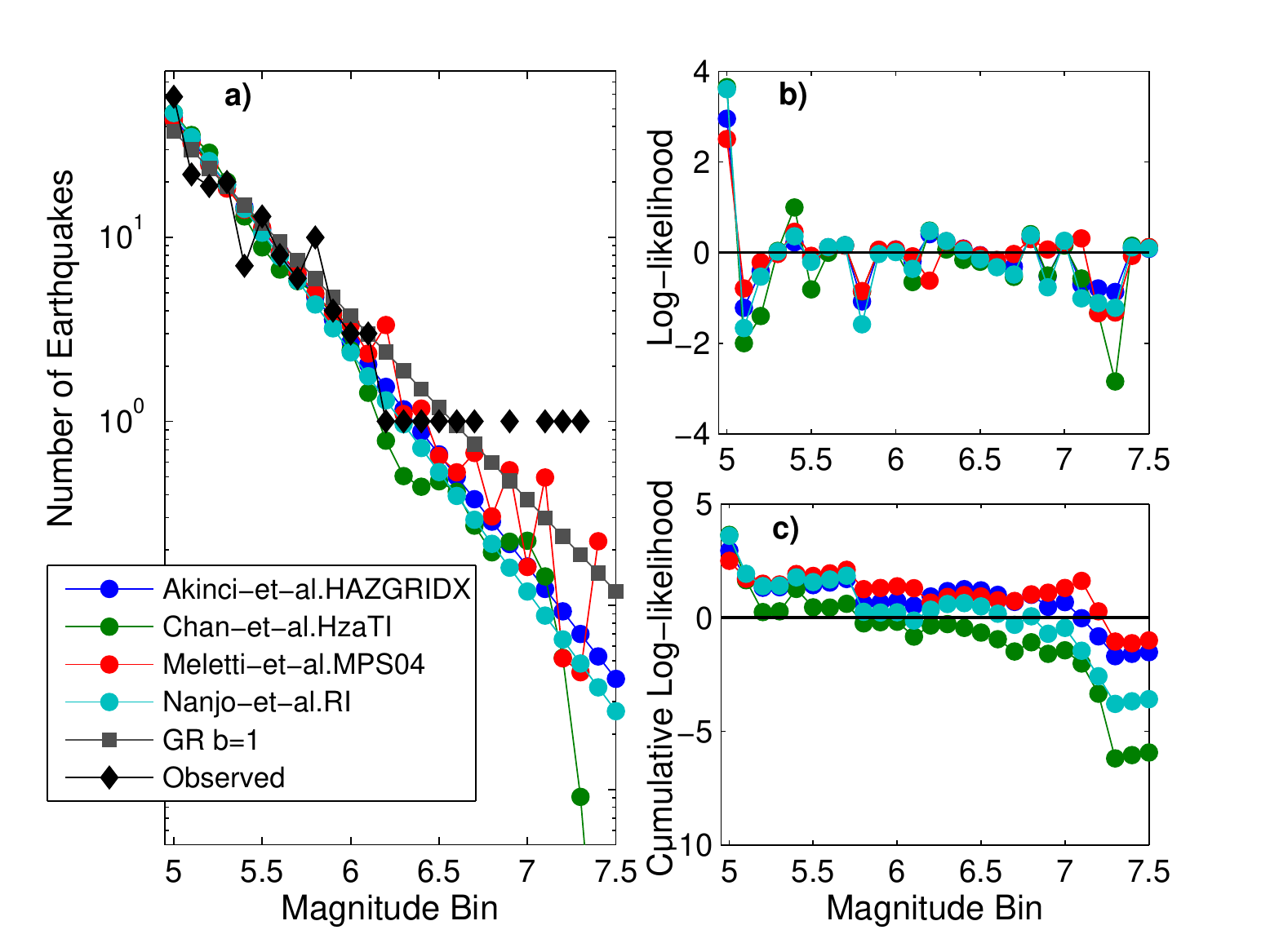}
\caption{\label{fig:m} Magnitude distributions and likelihood scores of the four models that failed the M-test on the 1901-2006 CPTI target period. a) observed and predicted histograms. b) bin-wise log-likelihood ratio of the models against a pure Gutenberg-Richter(GR) model with b-value equal to one. c) cumulative log-likelihood ratio.}
\end{figure}

Interestingly, four forecasts fail the M-test: \model{Akinci-et-al.Hazgridx}, \model{Chan-et-al.HzaTI}, \model{Meletti-et-al.MPS04} and \model{Nanjo-et-al.RI}. In Figure \ref{fig:m}, we compare the observed with their predicted magnitude distributions. For reference, we added a pure Gutenberg-Richter (GR) distribution with b-value equal to one, which passes the M-test. The magnitude distributions predicted by \model{Akinci-et-al.Hazgridx} and \model{Nanjo-et-al.RI} are close to exponential, but with b-values larger than one. As a result, large earthquakes are less likely, and the forecasts are penalized for the occurrence of three $m>7$ earthquakes. The magnitude distribution of the forecast  \model{Chan-et-al.HzaTI} seems to reflect its non-parametric kernel estimation method (see section \ref{sec:Models}) and also underpredicts the rate of large shocks. Finally, the magnitude distribution of \model{Meletti-et-al.MPS04} is non-monotonic: several characteristic magnitude bulges can be seen. However, the largest events occur between the bulges, for which the forecast is penalized.

\section{Discussion and Conclusions}
\label{sec:Disc}

\subsection{The Role of the Poisson Distribution in the Forecast Specification}

The assumption of Poisson rate variability in the CSEP-Italy experiments (as well as other CSEP experiments, including RELM \citep{Field2007, Schorlemmer-et-al2007}) has certain advantages.  In particular, this is a simplifying assumption: because the Poisson distribution is defined by a single parameter, the forecasts do not require a complete probability distribution in each bin.    Moreover, Poisson variability has often been used as a reference model against which to compare time-varying forecasts, and it yields an intuitive understanding.  

Despite these advantages, however, this assumption is questionable, and the method of forcing each forecast to be characterized by the same uncertainty is not the only solution \citep[see also the discussion by][]{Zechar-et-al2010}. \citet{WernerSornette2008a} remarked that most forecast models generate their own likelihood distribution, and this distribution depends on the particular assumptions of the model; moreover, there is no reason to force every model to use the same form of likelihood distribution.  The effect of this forcing is likely stronger for time-dependent, e.g. daily forecasts \citep{LombardiMarzocchi2010}, and it is difficult to judge (without the help of modelers) the quality of approximating each model-dependent distribution by a Poisson distribution. On the other hand, one can check whether or not the Poisson assumption is appropriate with respect to observations. In section \ref{sec:NBD}, we showed that the target earthquake rate distribution is approximated better by an NBD than by a Poisson distribution. Therefore, time-independent forecasts that predict a Poisson rate variability necessarily fail more often than expected at $95\%$ confidence because the observed distribution differs from the model distribution. To improve time-independent forecasts, the (non-Poissonian and potentially negative binomial) marginal rate distribution over long timescales needs to be estimated. However, the parameter values of the rate NBD change as a function of the temporal and spatial boundaries of the study region over available observational periods \citep{Kagan2009a}. Whether a stable asymptotic limit exists (loosely speaking, whether seismic rates are stationary) remains an open question. For time-dependent models, on the other hand, several classes exist which are capable of producing a rate NBD over finite time periods including branching processes \citep{Kagan2009a} and Poisson processes with a stochastic rate parameter distributed according to the Gamma distribution. 

Despite this criticism, it is unlikely that the Poisson distribution would be replaced by a model-dependent distribution that is substantially different, particularly for long-term models. Therefore, the p-values of the test statistics used in the N, L, S and M-tests might be biased towards lower values, but they do provide rough estimates. Nevertheless, one should bear in mind that a quantile score that is outside the $95\%$ confidence bounds of the Poisson distribution may be within the acceptable range if a model-dependent distribution were used. As an illustration, and to explore the effect of the Poisson assumption in these experiments, we created a set of modified forecasts with rate variability estimated from the observed history. The width of the $95\%$ confidence interval of the total rate forecast increased, in certain cases substantially. Several forecasts are rejected if a Poisson variability is assumed, while they pass the test under the assumption of an NBD. Overall, however, the p-values (quantile scores) of the test statistics based on the Poisson approximation often give good approximate values. Only in borderline cases did the Poisson assumption lead to (potentially) false rejections of forecasts. 

The modified forecasts based on an NBD are not an entirely satisfactory solution to the problem. First, the model distribution in each bin should arise naturally from a model's hypotheses, rather than an empirical adjustment made by those evaluating the forecast. Second, even if a negative binomial distribution adequately represents the distribution of the total number of observed events in an entire testing region, one should specify the parameter values for each bin to make the non-Poisson forecasts amenable also to the L-, S and M-tests. Therefore, future experiments should allow forecasts that are not characterized by Poisson rate uncertainty. 

More generally, future experiments might consider other forecast formats and additional model classes. For example, stochastic point process models provide a continuous likelihood function which can characterize conditional dependence in time, magnitude and space (and focal mechanisms, etc.). As a result, full likelihood-based inference for point processes and tools for model-diagnostics are applicable to this class of models \citep[e.g.][]{Ogata1999, DaleyVereJones2003, Schoenberg2003}. However, when considering new classes of forecasts, one may keep in mind that a major success of the RELM and CSEP experiments was the homogenization of forecast formats to facilitate comparative testing.

\subsection{Performance and Utility of the Tests}

We explored results from the N, L, S and M-tests in this study because they are the ``staple" CSEP tests. Other metrics for evaluating forecasts should certainly be considered, especially with regard to alarm-based tests \citep[e.g.][]{MolchanKeilisBorok2008, ZecharJordan2008} and further conditional likelihood tests \citep{Zechar-et-al2010}. Overall, the N, L, S and M-tests are intuitive and relatively easy to interpret. However, we demonstrated a weakness in the L-test and replaced it with a conditional L-test that better assesses the quality of a forecast \citep[see also][]{Werner-et-al2009b}. Among the metrics, the S-test results were the most helpful in tracking down the weak features of forecasts, because the biggest differences between time-independent models lie in their spatial forecasts.  

The M-test results were largely uninformative. Because the magnitude distributions considered here were so similar, this result is not surprising; indeed, it is in accordance with the statistical power exploration of \citet{Zechar-et-al2010}. No forecast could be rejected for target periods  ranging from 5 to 57 years. Different tests, such as a traditional Kolmogorov-Smirnov (KS) test, should be compared with the current likelihood-based M-test, particularly in terms of statistical power. 

The current status quo in CSEP experiments is to reject a forecast if it fails a single test at $95\%$ confidence. As we discussed above, the actual p-values provide a more meaningful assessment than a simple binary yes/no statement because the assumed confidence bounds may not accurately represent the model uncertainty. Furthermore, as the suite of tests grows, we should be concerned with joint confidence bounds of the ensemble of tests, rather than the individual significance levels of each test. Joint confidence bounds can be obtained from model simulations. A global confidence bound for the multiple tests can then be established. A similar question will arise when forecasts from the same model are tested within nested regions, as will be the case when considering the performance of a model's forecast for Italy with that for the entire globe. 

Finally, future experiments may consider developing tests that address particular characteristics of a forecast \citep[see also the discussion by][]{Zechar-et-al2010}. For example, a forecast might be a reflection of the hypothesis that the magnitude distribution varies as a function of tectonic setting. In this context, an M-test conditioned on the spatial distribution of observed earthquakes would provide a sharper test. 

\subsection{Overall Performance of the Forecasts}

A summary of all results can be found in Tables \ref{tab:sumCSI} and \ref{tab:sumCPTI}.  The Poisson N-test is possibly the strictest test within the present context, because none of the forecasts pass every N-test of the different periods. On the other hand, five forecasts pass all the N-tests with confidence bounds based on a negative binomial distribution (\model{Gulia-Wiemer.ALM}, \model{Gulia-Wiemer.HALM}, \model{Werner-et-al.CSI}, \model{Werner-et-al.Hybrid} and \model{Zechar-Jordan.CSI}). As we mentioned, several modelers indicated to us that their forecasts were calibrated on the moment magnitude scale. As a result, it is difficult to interpret their overpredictions beyond the obvious statement that the forecasts were poorly calibrated. The forecast \model{Nanjo-et-al.RI} is the only forecast that expects substantially fewer earthquakes than the observed sample mean, although the forecast fails the NBD N-test only for the longest of the considered target periods. Those forecasts that expect a number of shocks equal to the sample mean over their calibration period predict the number of quakes well, as should be expected. 

With one important exception, the conditional L-test results largely reflect the S-test results, because the predicted magnitude distributions were consistent with observations from all but the 106-year target period. The exception concerns the occurrence of an earthquake in a space-magnitude bin in which an earthquake should have been impossible according to the forecast: the 1968 $M_L=6.39$ Belice earthquake happened in a spatial cell in which the forecast \model{Meletti-et-al.MPS04} set a maximum magnitude of $M_L=6.25$. The discrepancy might be explained by the wrong magnitude conversion that the authors adopted and/or it may suggest that the model's assumptions regarding the spatial variation of maximum magnitudes may need to be revised. However, if we had tested the forecast against the moment magnitude of the Belice earthquake ($M_W6.33$, according to the CPTI catalog), the forecast would have still failed, thus pointing towards the latter explanation. 

The S-test results provided the most insight into the weaknesses of the forecasts. Only five forecasts pass all S-tests (\model{Akinci-et-al.Hazgridx}, \model{Chan-et-al.HzaTI}, \model{Werner-et-al.Hybrid}, \model{Zechar-Jordan.CPTI} and \model{Zechar-Jordan.Hybrid}). These forecasts fit the spatial distribution of the CSI and CPTI catalogs well, although they might overfit and perform poorly in the future. The models are also among the simplest, especially when compared to the forecast \model{Meletti-et-al.MPS04}. However, the forecasts \model{Werner-et-al.CSI} and \model{Zechar-Jordan.CSI}, which were calibrated on CSI data, cannot adequately forecast the spatial locations of earthquakes during the time period before the CSI data begins. This might indicate that the models are not smooth enough and do not anticipate sufficiently quiet regions becoming active. 

The ALM group of forecasts (\model{Gulia-Wiemer.ALM}, \model{Gulia-Wiemer.HALM} and \model{Schorlemmer-Wiemer.ALM}) consistently fail the S-tests, and often perform worse than a uniform forecast, because isolated earthquakes occur in extremely low-probability ``background" bins that cover roughly 50$\%$ of the region. We could not identify a common characteristic among the earthquakes that occurred in background bins.  The incurred likelihood losses cannot be compensated by the gains achieved by adequately forecasting the majority of earthquakes. The results suggest that the ALM forecasts are overly optimistic in ruling out earthquakes in their background bins, i.e. the models are not smooth enough. 

The forecast \model{Meletti-et-al.MPS04} also often fails the S-test because of a minority of earthquakes that occur in low-probability regions. Almost all earthquakes that incur likelihood losses are located offshore. But while the forecast performs substantially better onshore, a few surprising onshore earthquake locations remain. Poor performance of a forecast for offshore earthquakes potentially raises the problem of the ``weight" of each earthquake in the testing procedure. Specifically, if a model is intended for the practical purpose of seismic hazard assessment, then a rejection of its forecast due to offshore earthquakes may not have the same importance as a rejection due to earthquakes in regions of higher exposure and/or vulnerability.

Eight forecasts pass all M-tests (\model{Gulia-Wiemer.ALM}, \model{Gulia-Wiemer.HALM}, \model{Schorlemmer-Wiemer.ALM}, \model{Werner-et-al.CSI}, \model{Werner-et-al.Hybrid}, \model{Zechar-Jordan.CPTI}, \model{Zechar-Jordan.CSI} and \model{Zechar-Jordan.Hybrid}). Five of them are based on a simple Gutenberg-Richter distribution with uniform b-value equal to one (\model{Werner-et-al.CSI}, \model{Werner-et-al.Hybrid}, \model{Zechar-Jordan.CPTI}, \model{Zechar-Jordan.CSI} and \model{Zechar-Jordan.Hybrid}). This suggests that the hypothesis of a universally applicable, uniform Gutenberg-Richter distribution with b-value equal to one \citep[e.g.][]{BirdKagan2004} cannot be ruled out for the region of Italy. 

Four forecasts fail the M-test during the 1901-2007 target period of the CPTI catalog. The magnitude distributions of the forecasts  \model{Akinci-et-al.Hazgridx}, \model{Nanjo-et-al.RI}, \model{Chan-et-al.HzaTI} and \model{Meletti-et-al.MPS04} do not adequately forecast the largest magnitudes and the three observed $M_L>7$, in particular. In the case of the \model{Akinci-et-al.Hazgridx} and \model{Nanjo-et-al.RI} forecasts, the reason seems to be a b-value of the Gutenberg-Richter distribution that is too large. The non-parametric estimate of \model{Chan-et-al.HzaTI} also decays too quickly. The magnitude distribution of \model{Meletti-et-al.MPS04} reveals several characteristic magnitude values of elevated rates, but earthquakes also occur between them in extremely low-probability bins. However, these results should be interpreted cautiously because the same magnitude forecasts pass the 1950-2007 period, and because the greater uncertainty of the data prior to 1950 arguably influences the results.


\subsection{Value of Retrospective Evaluation}

The initial submission deadline for long-term earthquake forecasts for CSEP-Italy was 1 July 2009. Because the formal experiment was not intended to start until 1 August 2009, there was a brief period for initial analysis and quality control of the submitted forecasts. We provided a quick summary of the features of the forecasts and preliminary results of a retrospective evaluation to the modelers during this period. As a result, six of the eighteen time-independent and time-dependent long-term forecasts were modified and resubmitted before the final deadline on 1 August 2009. This initial quality control period was therefore extremely useful, and future experiments might consider expanding and formalizing the initial quality control period. 

The short one-month period was, however, too short to evaluate the forecasts retrospectively in the detail we present here. During the course of this study, the problem of the wrong magnitude scaling was discovered. Because at least 4 of the 18 forecasts are affected, a second round of submissions was solicited for 1 November 2009, and 15 revisions (and 2 new forecasts) were submitted. This suggests that the feedback provided to modelers based on the present study was useful and informative. The task of converting even a relatively simple hypothesis into a testable, probabilistic earthquake forecast should not be underestimated, and we suggest that future experiments include some form of retrospective testing prior to final submission.

The retrospective evaluation also showed that at least the time-independent forecasts can be evaluated in a meaningful manner and that useful information about the models can be extracted. 
Such information is critical for the development of better forecasts and for the evaluation of the underlying hypotheses of earthquake occurrence.



At the same time, retrospective evaluation cannot replace the prospective tests with zero degrees of freedom. Given the relative robustness of the results from the retrospective evaluation, we anticipate that the prospective experiment will provide further useful and more definite information about the quality of the forecasts. Most importantly, if the second round of forecast submissions contains significantly improved forecasts with fewer technical errors, we expect to see real progress in our understanding of earthquake predictability. 

\section*{Data and Sharing Resources}

We used two earthquake catalogs for this study: the {\it Catalogo Parametrico dei Terremoti Italiani} (Parametric Catalog of Italian Earthquakes, CPTI08) \citep{CPTI08} and the {\it Catalogo del la Sismicit`a Italiana} (Catalog of Italian Seismicity, CSI 1.1) \citep{Castello-et-al2007, Chiarabba-et-al2005}. The particular versions of the catalogs we used are available at \url{http://www.cseptesting.org/regions/italy}.

\section*{Acknowledgments}

We thank the following for their contribution to the CSEP-Italy Working Group: A. Akinci, C.H. Chan, A. Christophersen, R. Console, F. Euchner, L. Faenza, G. Falcone, M. Gerstenberger, L. Gulia, A.M. Lombardi, C. Meletti, M. Murru, K. Nanjo, B. Pace, L. Peruzza, D. Rhoades, D. Schorlemmer, M. Stucchi and J. Woessner. MJW was supported by the EXTREMES project of ETH's Competence Center Environment and Sustainability (CCES). Tests were performed within the European CSEP Testing Center at ETH Zurich, funded in parts through the European Union project NERIES. MJW thanks the Southern California Earthquake Center (SCEC) for travel support. SCEC is funded by NSF Cooperative Agreement EAR-0106924 and USGS Cooperative Agreement 02HQAG0008. The SCEC contribution number for this paper is 1436.




\begin{centering}
\section*{Author's Affiliations, Addresses}
\end{centering}

\noindent  Maximilian J. Werner, Swiss Seismological Service, Institute of Geophysics, ETH Zurich, Sonneggstr. 5, 8092 Zurich, Switzerland. 

\noindent J. Douglas Zechar, Swiss Seismological Service, Institute of Geophysics, ETH Zurich, Sonneggstr. 5, 8092 Zurich, Switzerland. 

\noindent Warner Marzocchi, Istituto Nazionale di Geofisica e Vulcanologia, Via di Vigna Murata, 605, 00143 Roma, Italy. 

\noindent  Stefan Wiemer, Swiss Seismological Service, Institute of Geophysics, ETH Zurich, Sonneggstr. 5, 8092 Zurich, Switzerland.

\newpage

\begin{sidewaystable}
\begin{center}
  \begin{tabular}{@{} llrrrl @{}}
    \toprule
      & & \multicolumn{1}{r}{Forecasted number} & Fraction of area & \\
      Model & Type & \multicolumn{1}{r}{of earthquakes} & covered by forecast (\%) & Reference \\
    \midrule
      \model{Akinci-et-al.Hazgridx}               & time-independent & 11.46    & 100.00  & \citet{Akinci2010} \\
      \model{Chan-et-al.HzaTI}     & time-independent & 14.76    & 100.00  & this issue \\
       \model{Gulia-Wiemer.ALM}  & time-independent & 8.58            & 100.00 & \citet{Gulia-et-al2010}  \\
       \model{Gulia-Wiemer.HALM}                 & time-independent & 9.53  & 100.00   & \citet{Gulia-et-al2010}  \\
      \model{Meletti-et-al.MPS04}        & time-independent & 15.60    & 100.00  & this issue \\
      \model{Nanjo-et-al.RI}         & time-independent & 2.78    & 100.00  & this issue \\
     \model{Schorlemmer-Wiemer.ALM}                 & time-independent & 12.74   & 100.00  & \citet{Gulia-et-al2010}  \\
       \model{Werner-et-al.CSI}              & time-independent& 6.21  & 100.00  & \citet{Werner-et-al2010c} \\
       \model{Werner-et-al.Hybrid}              & time-independent & 6.52   & 100.00  & \citet{Werner-et-al2010c} \\
     \model{Zechar-Jordan.CPTI}              & time-independent & 14.38    & 100.00  & \citet{ZecharJordan2010a} \\
       \model{Zechar-Jordan.CSI}              & time-independent & 5.88   & 100.00  & \citet{ZecharJordan2010a} \\
       \model{Zechar-Jordan.Hybrid}              & time-independent & 13.23    & 100.00  & \citet{ZecharJordan2010a} \\
    \midrule
       \model{Chan-et-al.HzaTD}           & time-dependent & 14.87          & 100.00 & this issue \\
       \model{Console-et-al.LTST}             & time-dependent & 10.98  & 100.00  & this issue  \\
      \model{Faenza-et-al.PHMgrid}    & time-dependent & 6.64  & 100.00  & this issue \\
      \model{Faenza-et-al.PHMzone} & time-dependent & 6.30           & 100.00 & this issue \\
      \model{Lombardi-Marzocchi.DBM}       & time-dependent & 9.06    & 100.00  & this issue  \\
      \model{Peruzza-et-al.LASSCI}        & time-dependent & 1.90    & 7.09  & this issue  \\
    \bottomrule
  \end{tabular}
\end{center}
\caption{Five- and ten-year CSEP-Italy forecasts being evaluated within the European CSEP Testing Center at ETH Zurich. Forecasts were submitted before 1 August 2009. The fraction of the area covered by forecast is the portion of the study region for which the model makes a forecast.}
\label{tab:models}
\end{sidewaystable}

\begin{table}
\begin{center}
  \begin{tabular}{l*{4}{c}}
    \toprule
      & \multicolumn{4}{c}{CSI}  \\
      Model & 1988-1992 & 1993-1997 & 1998-2002 & 1985-2003 \\
    \midrule
      \model{Akinci-et-al.Hazgridx} & N$^{+}$ & & & N$^{+}_{p}$ \\
      \model{Chan-et-al.HzaTI} & N$^{+}$ & & & N$^{+}$ \\
      \model{Gulia-Wiemer.ALM} & & $\hat{L}$, S & $\hat{L}$, S & $\hat{L}$, S \\
      \model{Gulia-Wiemer.HALM} & N$^{+}_{p}$ & $\hat{L}$, S & $\hat{L}$, S & $\hat{L}$, S \\
      \model{Meletti-et-al.MPS04} & N$^{+}$ & & $\hat{L}$, S & N$^{+}$, $\hat{L}$, S \\
      \model{Nanjo-et-al.RI} & & N$^{-}_{p}$ & N$^{-}_{p}$, L & N$^{-}_{p}$ \\
      \model{Schorlemmer-Wiemer.ALM} & N$^{+}$, $\hat{L}$, S & $\hat{L}$, S & $\hat{L}$, S &  N$^{+}$, $\hat{L}$, S \\
      \model{Werner-et-al.CSI} & & & & \\
      \model{Werner-et-al.Hybrid} & & & &  \\
      \model{Zechar-Jordan.CPTI} & N$^{+}$ & & & N$^{+}$ \\
      \model{Zechar-Jordan.CSI} & & & & \\
      \model{Zechar-Jordan.Hybrid} & N$^{+}$ & & & N$^{+}$ \\
    \bottomrule
  \end{tabular}
\end{center}
\caption{Summary results of the forecast tests obtained using the CSI catalog.  For each model and each experiment time period, the tests which the forecast failed are denoted, using a $5\%$ critical significance value.  For the N-test, N$^{+}$ indicates that the forecast overpredicted the observed rate, N- indicates an underprediction; the subscript $p$ indicates that the forecast only failed when assuming Poisson uncertainty, otherwise it failed under the Poisson and NBD. }
\label{tab:sumCSI}
\end{table}

\begin{table}
\begin{center}
  \begin{tabular}{l*{7}{c}}
    \toprule
      & \multicolumn{7}{c}{CPTI} \\
      Model & 57-66 & 67-76 & 77-86 & 87-96 & 97-06 & 1950-2006 & 1901-2006 \\
    \midrule
      \model{Akinci-et-al.Hazgridx} & & & N$^{+}_{p}$ & N$^{+}$ & &  N$^{+}$ & N$^{+}$ \\
      \model{Chan-et-al.HzaTI} & N$^{+}_{p}$ & N$^{+}_{p}$ & N$^{+}$ & N$^{+}$ & & N$^{+}$ & N$^{+}$\\
      \model{Gulia-Wiemer.ALM} &  $\hat{L}$, S & $\hat{L}$, S & & N$^{+}_{p}$ & S  & $\hat{L}$, S & $\hat{L}$, S\\
      \model{Gulia-Wiemer.HALM} & $\hat{L}$, S & $\hat{L}$, S & & N$^{+}_{p}$ & S & N$^{+}_{p}$, $\hat{L}$, S & $\hat{L}$, S \\
      \model{Meletti-et-al.MPS04} &  N$^{+}$, S & N$^{+}_{p}$, $\hat{L}$ & N$^{+}$ & N$^{+}$, $\hat{L}$, S & &  N$^{+}$, $\hat{L}$, S & N$^{+}$, $\hat{L}$, S \\
      \model{Nanjo-et-al.RI} & N$^{-}_{p}$ & N$^{-}_{p}$, $\hat{L}$, S & N$^{-}_{p}$ & & N$^{-}_{p}$ & N$^{-}$, S & N$^{-}$, $\hat{L}$, S \\
      \model{Schorlemmer-Wiemer.ALM} &  N$^{+}_{p}$, $\hat{L}$, S & $\hat{L}$, S & N$^{+}_{p}$, $\hat{L}$, S & N$^{+}$, $\hat{L}$, S & $\hat{L}$, S &  N$^{+}$, $\hat{L}$, S & N$^{+}$, $\hat{L}$, S\\
      \model{Werner-et-al.CSI} & S & $\hat{L}$, S & & & N$^{-}_{p}$ &  & N$^{-}_{p}$, $\hat{L}$, S\\
      \model{Werner-et-al.Hybrid} & & & & & N$^{-}_{p}$ &  & N$^{-}_{p}$\\
      \model{Zechar-Jordan.CPTI} & N$^{+}$ & N$^{+}_{p}$ & N$^{+}$ & N$^{+}$ & & N$^{+}$ & N$^{+}$\\
      \model{Zechar-Jordan.CSI} & S & S & & & N$^{-}_{p}$ &  & N$^{-}_{p}$, S \\
      \model{Zechar-Jordan.Hybrid} & N$^{+}_{p}$ & & N$^{+}$ & N$^{+}$ & &  N$^{+}$ & N$^{+}$ \\
    \bottomrule
  \end{tabular}
\end{center}
\caption{Summary results of the ten-year forecast tests obtained using the CPTI catalog.  For each model and each experiment time period, the tests which the forecast failed are denoted, using a 5\% critical significance value.  For the N-test, N$^{+}$ indicates that the forecast overpredicted the observed rate, N$^{-}$ indicates an underprediction; the subscript $p$ indicates that the forecast only failed when assuming Poisson uncertainty, otherwise it failed under the Poisson and NBD. }
\label{tab:sumCPTI}
\end{table}

\newpage
\clearpage

\newpage
\clearpage


\section*{Appendix: Negative-Binomial Forecasts}
\label{sec:AppA}

To create NBD forecasts, we used each forecast's total expected rate as the average of the distribution, and we fixed the variance of the forecast to be equal to  the observed sample variance from the CPTI catalog (estimated in section \ref{sec:NBD}). Thus, for five-year experiments, we use $\sigma^2_{5yr} = 23.73$, while for ten-year experiments, we use  $\sigma^2_{10yr} = 64.54$. 

For longer time periods (e.g., the durations of the CSI and CPTI catalogs), for which we cannot estimate directly the sample variance, we used the property that the variance of a finite sum of uncorrelated random variables is equal to the sum of their variances. We treated the numbers of observed earthquakes as uncorrelated random variables, meaning that we assumed that the numbers of observed earthquakes in adjacent time intervals are independent of each other. This is likely to be a better approximation for the ten-year intervals.  We computed the variance $\sigma^2(T)$ over some finite interval of $T$ years from the reference variance $\sigma^2_{10yr}$ using
\begin{equation}
\sigma^2 (T) = \frac{T}{10} \ \sigma^2_{10yr}
\label{eq:sigma}
\end{equation}

Table \ref{tab:NBD} lists the estimated and calculated variances for the various time intervals we used in this study. The NBD parameters, if needed, can be estimated from equations (\ref{eq:NBDAvg}) and (\ref{eq:NBDVar}). Because the direct estimate of $\sigma^2_{10yr}$ is larger than twice $\sigma^2_{5yr}$, it seems that there may be correlations at the five-year time scale. Alternatively, the sample size may be too small, because the  $95\%$ confidence intervals are large. 

\begin{table}[htbp]
\caption{ {\bf Estimated variances of the numbers of observed earthquakes for different time intervals:} \newline
*The variance was estimated directly from the catalog. Others were computed using equation (\ref{eq:sigma}).}
  \centering
  \begin{tabular}{@{} |c|c| @{}}
    \hline
Time Interval $T$ [yrs] & Estimated $\sigma^2(T)$ \\
    \hline
5 & 23.73* \\
10 & 64.54* \\
18 & 116.17 \\
57 &  367.88 \\
106 & 684.12 \\
    \hline
   \end{tabular}
  \label{tab:NBD}
\end{table}

\end{document}